\documentclass[12pt]{article}
\usepackage{}
\usepackage{amssymb}
\usepackage{amsmath}
\usepackage{graphicx}
\usepackage{indentfirst}
\usepackage{cite}

\begin{document}

\title{Temperature-dependent cross sections \\
for charmonium dissociation in collisions with \\
pions and rhos in hadronic matter}
\author{Jie Zhou \and Xiao-Ming Xu}
\date{}
\maketitle \vspace{-1cm}
\centerline{Department of Physics, Shanghai
University, Baoshan, Shanghai 200444, China}
\begin{abstract}
Meson-charmonium dissociation reactions governed by the quark interchange are 
studied with temperature-dependent quark potentials. Quark-antiquark 
relative-motion wave functions and masses of charmonia and charmed mesons are 
determined by the central spin-independent part of the potentials or by the
central spin-independent part and a smeared spin-spin interaction. The
prominent 
temperature dependence of the masses is found. Based on the potentials, the 
wave functions, and the meson masses, we obtain temperature-dependent cross 
sections for the fifteen dissociation reactions: $\pi J/\psi \to \bar{D}^{*} D$
or $\bar D D^*$, $\pi J/\psi \to\bar{D}^{*} D^{*}$, 
$\pi {\psi}' \to \bar{D}^* D$ or $\bar D D^*$, $\pi {\psi}' \to \bar{D}^* D^*$,
$\pi \chi_{c} \to \bar{D}^* D$ or $\bar D D^*$, 
$\pi \chi_{c} \to \bar{D}^* D^*$, $\rho J/\psi \to \bar{D} D$, 
$\rho J/\psi \to \bar{D}^* D$ or $\bar D D^*$, 
$\rho J/\psi \to \bar{D}^* D^*$, $\rho {\psi}' \to \bar{D} D$, 
$\rho {\psi}' \to \bar{D}^* D$ or $\bar D D^*$, 
$\rho {\psi}' \to \bar{D}^* D^*$, $\rho \chi_{c} \to \bar{D} D$, 
$\rho \chi_{c} \to \bar{D}^* D$ or $\bar D D^*$, and 
$\rho \chi_{c} \to \bar{D}^* D^*$. The numerical cross sections are 
parametrized for future applications in hadronic matter. The particular 
temperature dependence of the $J/\psi$ bound state leads to unusual behavior of
the cross sections for endothermic $J/\psi$ dissociation reactions. The quantum
numbers of ${\psi}'$ and $\chi_c$ can not make their difference in mass in the 
temperature region $0.6T_{\rm c} \leq T < T_{\rm c}$, but can make the 
${\psi}'$ dissociation different from the $\chi_c$ dissociation.
\end{abstract}


\newpage

\leftline{\bf 1. Introduction}
\vspace{0.5cm}

From the year when Matsui and Satz \cite{MS} originally suggested the 
suppressed ${J/\psi}$ production as a signature for the formation of a 
quark-gluon plasma in high-energy heavy-ion collisions to the last year's 
quark 
matter conference where preliminary $J/\psi$ data in Pb-Pb collisions at the 
Large Hadron Collider were reported by the ALICE Collaboration \cite{AT} and 
the CMS Collaboration \cite{BW}, important measurements have been done. 
Essential theoretical progress on $J/\psi$ has also been made. One of the 
fundamental issues is the dissociation of charmonia in hadronic matter 
\cite{FLP}. To identify $J/\psi$ as a probe of the quark-gluon plasma in a 
definite way, hadron-charmonium dissociation processes must be well understood.
Calculations of dissociation cross sections are thus an important aspect in 
studying $J/\psi$ physics. The dissociation is described by the meson or quark 
degree of freedom. Corresponding to the two degrees of freedom, different 
scattering mechanisms can be assumed, and different results on the dissociation
cross sections have been reported in the literature.

There are mainly three approaches to the meson-$J/\psi$ dissociation problem. 
In the short-distance approach the parton model of light hadrons, the 
constituent quark model of $J/\psi$, and the gluon-$J/\psi$ dissociation cross 
section of Peskin and Bhanot \cite{MEP} are employed in Refs. \cite{KS,AGGA} to
investigate $\pi$$J/\psi$ and $N$$J/\psi$ dissociation. The portion of hard 
gluons inside the nucleon or pion at low energies is not large enough to induce
mb-scale cross sections. In the meson-exchange approach $J/\psi$ dissociation 
has been studied with effective meson Lagrangians. Since the discovery of 
$J/\psi$ reveals the charm quark, $J/\psi$ is first described by QCD. The meson
field description of $J/\psi$ and charmed mesons began to appear from the work 
of Matinyan and M\"uller \cite{MM} who first discussed the t-channel exchange 
of $D$ meson in inelastic $\pi J/\psi$ and $\rho J/\psi$ scattering. They got 
mb-scale cross sections for $\pi J/\psi \to D^*\bar{D} + D \bar D^*$ and $\rho 
J/\psi \to D\bar{D}$ at low energies. A similar scale of cross sections has 
been obtained in other meson Lagrangians with different symmetries and modified
vertex functions in Feynman diagrams to include the effect of finite meson form
factors \cite{LK,KLH,OSL,NNR,MPPR,BG}. In the quark-interchange approach 
$J/\psi$ dissociation has been studied in nonrelativistic quark potential 
models \cite{MBQ,WSB,WSB01,BSWX}. The assumption that color-independent 
confining interaction acts only between a quark and an antiquark in Ref. 
\cite{MBQ} 
yields that the ${J/\psi}$ absorption cross section corresponding to $\pi 
J/\psi \to D^* \bar D + D \bar D^* + D^* \bar D^*$ has a peak value of about 7 
mb at the kinetic energy $E_{\rm kin} \equiv \sqrt{s}-(m_{\pi}
+m_{J/\psi})\simeq0.8$ GeV. However, by including color generators in the 
linear confining potential and allowing the interaction to connect any two 
constituents (quarks and antiquarks), Wong, Swanson, and Barnes \cite{WSB} gave
a peak value of only $\sim 1$ mb at the same kinetic energy. Usefully, the 
dissociation cross sections of ground state, orbitally and radially excited 
charmonia in collisions with $\pi$ and $\rho$ mesons have been presented in 
Ref. \cite{BSWX} where all parameters in the color Coulomb, spin-spin hyperfine
and linear confining interactions are determined by fits to the experimental 
meson spectrum.

All the above works only involve charmonium dissociation reactions in vacuum, 
and have been used to evaluate $J/\psi$ suppression in nucleus-nucleus 
collisions, but we know that hadron masses, the quark potential and so on are 
affected by medium and the QCD phase transition is the most striking medium 
effect, so we must study medium effects on the charmonium dissociation. In 
this work we calculate dissociation cross sections of $J/\psi$, ${\psi}'$, and 
$\chi_c$ in collisions with $\pi$ and $\rho$ at various temperatures. The 
quark-interchange mechanism \cite{BS92}, the Born approximation, and a 
temperature-dependent quark potential \cite{ZXG} are ingredients in 
establishing cross section formulas.

This paper is organized as follows. In the next section we use the 
temperature-dependent potential \cite{ZXG}
in the Schr\"odinger equation to obtain 
temperature-dependent masses of charmonia and charmed mesons. In Section 3 we 
give formulas for charmonium dissociation cross sections. Numerical results for
unpolarized cross sections of $J/\psi$, ${\psi}'$, and $\chi_c$ in collisions 
with $\pi$ and $\rho$ at six temperatures are shown in Section 4 and relevant 
discussions are given. In Section 5 the spin-spin interaction arising from 
one-gluon exchange is smeared. Subsequent results are shown and numerical cross
sections are parametrized. In 
Section 6 we present a procedure on how to obtain unpolarized cross sections at
any temperature in the region $0.65 \leq T/T_{\rm c} < 1$ where $T_{\rm c}$ is 
the critical temperature. Finally, we summarize the present work in Section 7.

\vspace{0.5cm}
\leftline{\bf 2. Masses of charmonia and charmed mesons}
\vspace{0.5cm}

As shown in Refs. \cite{DGG,IS1,GI,CI}, the quark potential containing 
flavor-independent confinement, a Coulomb term and hyperfine interactions can 
consistently reproduce masses from light to heavy hadrons, and the 
flavor-independent assumption of confinement is thus reasonable. Such 
confinement in hadronic matter can be estimated by the lattice calculations 
\cite{KLP01} and depends on temperature. At large distances the confinement 
manifests itself by a plateau that lowers with increasing temperature. In Ref. 
\cite{ZXG} we used the confinement at large distances and the short-distance 
potential originating from one-gluon exchange and loop corrections in 
perturbative QCD \cite {BT} to construct a central spin-independent, 
flavor-independent but temperature-dependent potential
\begin{equation}
V_{\rm {si}}(\vec {r})=-\frac {\vec {\lambda}_a}{2} \cdot \frac {\vec 
{\lambda}_b}{2}
\frac {3}{4} D \left[ 1.3- \left( \frac {T}{T_{\rm c}} \right)^4 \right]
\tanh (Ar) + \frac {\vec {\lambda}_a}{2} \cdot \frac {\vec {\lambda}_b}{2}
\frac {6\pi}{25} \frac {v(\lambda r)}{r} \exp (-Er)  .
\end{equation}
Only here $D=0.7~{\rm GeV}$, $T_{\rm c}=0.175~{\rm GeV}$, 
$A=1.5[0.75+0.25(\frac {T}{T_{\rm c}})^{10}]^6~{\rm GeV}$, $E=0.6~{\rm GeV}$, 
and $\lambda=\sqrt {3b_0/16\pi^2\alpha'}$ in which 
$\alpha'=1.04~{\rm GeV}^{-2}$ is the Regge slope and $b_0=11-\frac{2}{3} N_f$
with the quark flavor number $N_f=4$. $\vec \lambda_a$ are the Gell-Mann 
matrices for the color generators of constituent $a$.
The dimensionless function $v(x)$ is an integration over the absolute value of 
gluon momentum $\vec {Q}$
\begin{equation}
v(x)=\frac
{4b_0}{\pi} \int^\infty_0 \frac {dQ}{Q} (\rho (\vec {Q}^2) -\frac {K}{\vec
{Q}^2}) \sin (\frac {Q}{\lambda}x) ,
\end{equation}
where $K= 3/16\pi^2\alpha'$. The subtraction of $ {K}/{\vec{Q}^2}$ from the 
physical running coupling constant $\rho (\vec {Q}^2)$ leaves only the 
contribution of one-gluon exchange plus perturbative one- and two-loop 
corrections. The factor $\exp (-Er)$ is a medium modification factor to the 
potential of one-gluon exchange plus perturbative one- and two-loop 
corrections. The temperature dependence is completely negligible at very short 
distances and obvious at intermediate and large distances. The potential well 
fits the lattice gauge results at $T/T_{\rm c} > 0.55$ \cite{KLP01}, but fails 
to give a long-ranged linear confining potential at $T=0$.

Given the charm quark mass $m_c=1.51$ GeV, the Schr\"odinger equation with the 
potential given in Eq. (1) is solved to obtain masses and wave functions of 
$J/\psi$, ${\psi}'$, and $\chi_c$. Here the $\chi_c$ mass corresponds to the 
center of gravity of $\chi_{c0}$, $\chi_{c1}$, and $\chi_{c2}$ \cite {EPRD}. At
$T=0$ the masses of $J/\psi$, ${\psi}'$, and $\chi_c$ are 3.10505 GeV, 3.67679 
GeV, and 3.51138 GeV, compared to the experimental values 3.096916 GeV, 3.68609
GeV, and 3.5253 GeV \cite {KN2010}, respectively. The temperature dependence of
the charmonium masses is shown by the solid, dashed and dotted curves in Fig. 
1. Because $J/\psi$ has a small radius, the $J/\psi$ mass changes more slowly 
than the ${\psi}'$ and $\chi_c$ masses. Even though ${\psi}'$ and $\chi_c$ have
different quantum numbers, both become degenerate in mass in the temperature 
region in Fig. 1. Furthermore, at temperatures very close to the critical 
temperature $J/\psi$ joins ${\psi}'$ and $\chi_c$ to form a triplet in mass. 
This is a medium effect on charmonia!

Given $m_c=1.51$ GeV, the up and down quark masses $m_u=m_d=$0.32 GeV, and the 
strange quark mass $m_s=$0.5 GeV, the Schr\"odinger equation with the central 
spin-independent potential offers the same quark-antiquark relative-motion wave
functions of  $\pi$ and $\rho$ ($D$ and $D^*$, $D_s$ and $D^*_s$) and the 
spin-averaged mass of $\pi$ and $\rho$ ($D$ and $D^*$, $D_s$ and $D^*_s$). The 
spin-averaged mass of a spin-0 meson and a spin-1 meson with the same isospin 
is one-fourth of the spin-0 meson mass plus three-fourths of the spin-1 meson 
mass. The quark-antiquark relative-motion wave functions of $\pi$ and $\rho$ 
($D$ and $D^*$, $D_s$ and $D^*_s$) are used to calculate the mass splitting of 
$\pi$ and $\rho$ ($D$ and $D^*$, $D_s$ and $D^*_s$) with the spin-spin 
interaction that arises from one-gluon exchange plus one- and two-loop 
corrections \cite{XMX}
\begin{equation}
V_{\rm ss}=
- \frac {\vec {\lambda}_a}{2} \cdot \frac {\vec {\lambda}_b}{2}
\frac {16\pi^2}{25} \delta^3(\vec {r})  \frac {\vec {s}_a \cdot \vec
{s}_b}{m_am_b}
+ \frac {\vec {\lambda}_a}{2} \cdot \frac {\vec {\lambda}_b}{2}
  \frac {4\pi}{25} \frac {1}{r}
\frac {d^2v(\lambda r)}{dr^2} \frac {\vec {s}_a \cdot \vec {s}_b}{m_am_b} ,
\end{equation}
where $\vec {s}_a$ ($\vec {s}_b$) and $m_a$ ($m_b$) are the spin and mass of 
the constituent $a$ ($b$), respectively. From the mass splitting and the 
spin-averaged mass we get the spin-0 meson mass and the spin-1 meson mass: the 
former is the spin-averaged mass minus three-fourths of the mass splitting; the
latter is the spin-averaged mass plus one-fourth of the mass splitting. At 
$T=0$ the masses of $D$, $D^*$, $D_s$, and $D^*_s$ are 1.80666 GeV, 2.09552 
GeV, 1.85228 GeV, and 2.18695 GeV, compared to the measured values 1.86722 GeV,
2.00861 GeV, 1.96847 GeV, and 2.1123 GeV, respectively. The temperature 
dependence of the $D$, $D^*$, $D_s$ and $D^*_s$ masses is plotted in Fig. 1 as 
two long dashed curves and two dot-dashed curves. The $D$ and $D_s$ masses keep
almost unchanged from $T=0.6T_{\rm c}$ to $0.85T_{\rm c}$ and $0.9T_{\rm c}$, 
respectively, and apparently fall off in the other temperature regions. The 
$D^*$ and $D^*_s$ masses decrease slowly from $T=0.6T_{\rm c}$, and become 
apparently falling off from $T=0.8T_{\rm c}$ and $T=0.85T_{\rm c}$, 
respectively. The medium effect on charmed mesons is obvious only in the region
where the masses apparently fall off. The temperature dependence of $\pi$ and 
$\rho$ masses was shown in Fig. 2 of Ref. \cite{ZXG}. With increasing 
temperature the $\pi$ mass decreases slowly for $0.6T_{\rm c} \leq T < 
0.78T_{\rm c}$ and rapidly for $T \geq 0.78T_{\rm c}$ while the $\rho$ mass 
decreases rapidly for $0.6T_{\rm c} \leq T <T_{\rm c}$. From $T=0.6 T_{\rm c}$ 
to $0.99 T_{\rm c}$ the masses of $\pi$, $\rho$, $J/\psi$, ${\psi}'$, $\chi_c$,
$D$, $D^*$, $D_s$, and $D^*_s$ are reduced by $100 \%$, $99 \%$, $7 \%$, 
$16 \%$, $16 \%$, $20 \%$, $27 \%$, $12 \%$, and $23 \%$, respectively. 
Therefore, the medium effect on the two light mesons is more obvious. Either 
$D$ and $D^*$ or $D_s$ and $D^*_s$ become a doublet in mass at $T \to 
T_{\rm c}$, but the four mesons do not become degenerate unlike $J/\psi$, 
${\psi}'$ and $\chi_c$.

The meson masses in units of GeV in the region $0.6 \leq T/T_{\rm c} < 1$ are 
parametrized as
\begin{equation}
m_{J/\psi}=3.07 \left[ 1-\left( \frac{T}{1.01T_{\rm c}} \right)^{3.76} 
\right]^{0.03} ,
\end{equation}
\begin{equation}
m_{{\psi}'}=3.48 \left[ 1-\left( \frac{T}{2.19T_{\rm c}} \right)^{4.63} 
\right]^{7.74} ,
\end{equation}
\begin{equation}
m_{\chi_c}=3.42 \left[ 1-\left( \frac{T}{1.89T_{\rm c}} \right)^{5.65} 
\right]^{6.96} ,
\end{equation}
\begin{equation}
m_{D}=1.795 \left[ 1-\left( \frac{T}{1.16T_{\rm c}} \right)^{9.67} 
\right]^{0.92} ,
\end{equation}
\begin{equation}
m_{D^*}=2.02 \left[ 1-\left( \frac{T}{1.42T_{\rm c}} \right)^{5.38} 
\right]^{2.18} ,
\end{equation}
\begin{equation}
m_{D_s}=1.94 \left[ 1-\left( \frac{T}{1.02T_{\rm c}} \right)^{3.3} 
\right]^{0.08},
\end{equation}
\begin{equation}
m_{D^*_s}=2.133 \left[ 1-\left( \frac{T}{1.29T_{\rm c}} \right)^{6.28} 
\right]^{1.29} .
\end{equation}

\vspace{0.5cm}
\leftline{\bf 3. Cross section formulas }
\vspace{0.5cm}

In nonrelativistic dynamics for the quark-interchange process $q\bar q + c \bar
c \to q \bar c + c \bar q$, the center-of-mass motion of $q \bar q$ and $c \bar
c$ (i.e. $q \bar c$ and $c \bar q$) is separated off. This guarantees that 
cross sections are calculated in a way independent of the center of mass. We 
then choose the center-of-mass frame where the cross section can be easily 
formulated \cite {LX}. We denote the mass and the four-momentum of meson $i$ 
$(i=q\bar q,$ $c \bar c,$ $q \bar c,$ $c \bar q)$ by $m_i$ and $P_i=(E_i,\vec 
{ P_i})$, respectively. The Mandelstam variables are $s=(E_{q\bar q}+E_{c \bar 
c})^2-(\vec {P}_{q\bar q}+\vec {P}_{c \bar c})^2$ and
$t=(E_{q\bar q}-E_{q \bar c})^2-(\vec {P}_{q\bar q}-\vec {P}_{q \bar c})^2$. 
The cross section for the meson-charmonium scattering $q\bar q + c \bar c \to 
q \bar c + c \bar q$ in the center-of-mass frame is
\begin{equation}
\sigma(S,m_S,\sqrt {s},T) =\frac{1}{32\pi s}\frac{|\vec{P}^{\prime }(\sqrt{s})|
}{|\vec{P}(\sqrt{s})|}\int_{0}^{\pi }d\theta
|\mathcal{M}_{\rm fi} (s,t)|^{2}\sin \theta ,
\end{equation}
where $S$ is the total spin of either the two incoming mesons or the two 
outgoing mesons, $m_S$ is the magnetic projection quantum number of $S$, 
${\cal M}_{\rm fi}$ is the transition amplitude, and $\theta$ is the angle 
between $q \bar q$ momentum $\vec{P}$ and $q \bar c$ momentum 
$\vec{P}^{\prime}$. $\vec{P}$ and $\vec{P}^{\prime}$ are related to the 
Mandelstam variable $s$ by
\begin{equation}
\vec {P}^2(\sqrt{s})=\frac{1}{4s}\left\{ \left[ s-\left(
m_{q\bar q}^{2}+m_{c\bar c}^{2}\right) \right]^{2}
-4m_{q \bar q}^{2}m_{c\bar c}^{2} \right\},
\end{equation}
\begin{equation}
\vec {P'}^{2}(\sqrt{s})=\frac{1}{4s}\left\{ \left[
s-\left( m_{q\bar c}^2+m_{c \bar{q}}^2\right)\right]^2
-4m_{q \bar c}^2m_{c \bar{q}}^2 \right\} .
\end{equation}

The interchange of quarks brings about two scattering forms, the prior form and
the post form. The two forms may lead to different values of the transition 
amplitude ${\cal M}_{\rm fi}$ (and the subsequent cross section), which is the 
so-called post-prior discrepancy\cite{MM65,BBS01,WC}. Scattering in the prior 
form means that gluon exchange takes place prior to the quark interchange, and 
the corresponding transition amplitude is
\begin{eqnarray}
{\cal M}_{\rm fi}^{\rm prior} & = &
4\sqrt {E_{q \bar q} E_{c \bar c} E_{q \bar c} E_{c \bar q}}
<\psi_{q\bar {c}}|<\psi_{c \bar {q}}|(V_{q\bar {c}}+V_{c \bar {q}}+V_{qc}
+V_{\bar {q}\bar {c}})|\psi_{q\bar {q}}>|\psi_{c\bar {c}}>,     \nonumber   \\
\end{eqnarray}
while scattering in the post form means that the quark interchange is followed 
by gluon exchange, and the corresponding transition amplitude is
\begin{eqnarray}
{\cal M}_{\rm fi}^{\rm post} & = &
4\sqrt {E_{q \bar q} E_{c \bar c} E_{q \bar c} E_{c \bar q}}
<\psi_{q\bar {c}}|<\psi_{c \bar {q}}|(V_{q\bar {q}}+V_{c \bar {c}}+V_{qc}
+V_{\bar {q}\bar {c}})|\psi_{q\bar {q}}>|\psi_{c\bar {c}}>,     \nonumber   \\
\end{eqnarray}
where $\psi_{q\bar {c}}$ is the product of color, spin, flavor and 
momentum-space wave functions of the relative motion of $q$ and $\bar{c}$ and 
satisfies $\int \frac {d^3p_{q \bar{c}}}{(2\pi)^3} \psi^+_{q \bar{c} } 
(\vec {p}_{q \bar{c}})\psi_{q \bar{c}} (\vec {p}_{q \bar{c}}) =1$ where $\vec 
{p}_{q \bar{c}}$ is the relative momentum of $q$ and $\bar c$, and similarly 
$\psi_{c\bar {q}}$, $\psi_{q\bar {q}}$ and $\psi_{c\bar {c}}$. A relative 
momentum depends on a linear combination of $\vec P$ and $\vec P'$. The 
momentum-space wave functions are the Fourier transform of the coordinate-space
wave functions which are solutions of the Schr\"odinger equation. 
In the transition amplitude 
we use the Fourier transform of the sum of the central spin-independent 
potential and the spin-spin interaction:
\begin{eqnarray}
V_{ab}\left( \vec {Q}\right) &=& -\frac{ \vec {\lambda }_{a}}{2}
\cdot \frac{\vec {\lambda }_{b}}{2}\frac{3}{4} D
\left[ 1.3- \left( \frac {T}{T_{\rm c}} \right)^4 \right]
\left[ (2\pi)^3\delta^3 (\vec {Q}) - \frac {8\pi}{Q}
\int^\infty_0 dr \frac {r\sin (Qr)}{\exp (2Ar)+1} \right]
                               \notag \\
& &
+\frac{ \vec {\lambda }_{a}}{2} \cdot \frac{\vec {\lambda }_{b}}{2} 64 \pi E
\int^\infty_0 dq \frac {\rho (q^2) -\frac {K}{q^2}}{(E^2+Q^2+q^2)^2-4Q^2q^2}
                               \notag \\
& & -\frac{\vec {\lambda }_{a}}{2}
\cdot \frac{\vec {\lambda }_{b}}{2}\frac{16\pi ^{2}}{25}\frac{
\vec {s}_{a}\cdot \vec {s}_{b}}{m_{a}m_{b}}
+\frac{\vec {\lambda }_{a}}{2}\cdot \frac{\vec {\lambda }_{b}}{
2}\frac{16\pi ^{2}\lambda }{25Q}\int_{0}^{\infty}dx\frac{d^{2}v\left(
x\right) }{dx^{2}}\sin \left( \frac{Q}{\lambda }x\right)
\frac{\vec {s}_{a}\cdot \vec {s}_{b}}{m_{a}m_{b}} .
                                 \notag \\
\end{eqnarray}
Let $\sigma^{\rm prior}$ and $\sigma^{\rm post}$ be the cross sections for 
scattering in the prior form and in the post form, respectively, and they are 
given by Eq. (11). The unpolarized cross section for $q\bar q + c \bar c \to q 
\bar c + c \bar q$ is
\begin{eqnarray}
\sigma ^{\rm unpol} (\sqrt {s},T) & = &
\frac{1}{(2S_{q \bar q}+1)(2S_{c \bar c}+1)}
                                  \notag   \\
& & \times \sum _{S}(2S+1) \frac{\sigma^{\rm prior}(S,m_S,\sqrt {s},T)
+\sigma^{\rm post}(S,m_S,\sqrt {s},T)}{2} ,
                                  \notag   \\
\end{eqnarray}
where $S_{q \bar q}$ and $S_{c \bar c}$ are the spins of $q \bar q$ and 
$c \bar c$, respectively.

\vspace{0.5cm}
\leftline {\bf 4. Numerical cross sections and discussions}
\vspace{0.5cm}

Quark-antiquark relative-motion wave functions of mesons are given by the
Schr\"odinger equation 
with the central spin-independent potential $V_{\rm {si}}$.
With the up (down) quark mass 0.32GeV and the experimental $\pi$ mass, the 
experimental data of $S$-wave $I=2$ elastic phase shifts for $\pi \pi$ 
scattering in vacuum \cite {EC,NBD,WH,MJL} for $0 < \sqrt {s} < 2.4 $ GeV are 
reproduced from the potential in Eq. (16) at $T=0$ \cite {ZXG}. In the 
estimates of charmonium dissociation cross sections at $T=0$, the experimental 
masses of pion, rho, charmonia, and charmed mesons are employed. For $0.6 < 
T/T_{\rm c} < 1$ we use the temperature-dependent meson masses shown in Fig. 1 
and the $\pi$ and $\rho$ masses in Fig. 2 of Ref. \cite {ZXG}. Temperature 
dependence of charmonium dissociation cross sections shown in the next 
subsection seems to be complicated, but is understandable.

From $T=0.6 T_{\rm c}$ to $0.99 T_{\rm c}$ the masses of  $\pi$, $\rho$, 
$J/\psi$, ${\psi}'$, $\chi_c$, $D$, and $D^*$ are reduced by 0.40955 GeV, 
0.60936 GeV, 0.21908 GeV, 0.56022 GeV, 0.53639 GeV, 0.35318 GeV, and 0.53845 
GeV, respectively. The reduced amounts of ${\psi}'$, $\chi_c$ and $D^*$ masses 
are between the ones of $\pi$ and $\rho$ masses. Hence, the difference 
$m_{q \bar c} + m_{c \bar q} - m_{q \bar q} - m_{c \bar c}$ may be larger or 
smaller than zero. Whether a reaction is endothermic or not depends on 
temperature. A reaction may be exothermic below a temperature and endothermic 
above the temperature.

\vspace{0.5cm}
\leftline {\sl 4.1. Numerical cross sections}
\vspace{0.5cm}

The temperature-dependent potential given in Eq. (16), the wave functions, and 
the meson masses in the transition amplitude make both the meson-charmonium 
dissociation cross sections and relevant threshold energies dependent on 
temperature. In Figs. 2-16 we plot cross sections for the following fifteen 
meson-charmonium dissociation reactions: $\pi J/\psi \to \bar{D}^{*} D$ or 
$\bar D D^*$, $\pi J/\psi \to\bar{D}^{*} D^{*}$, $\pi {\psi}' \to \bar{D}^* D$ 
or $\bar D D^*$, $\pi {\psi}' \to \bar{D}^* D^*$, $\pi \chi_{c} \to 
\bar{D}^* D$ or $\bar D D^*$, $\pi \chi_{c} \to \bar{D}^* D^*$, $\rho J/\psi 
\to \bar{D} D$, $\rho J/\psi \to \bar{D}^* D$ or $\bar D D^*$, $\rho J/\psi \to
\bar{D}^* D^*$, $\rho {\psi}' \to \bar{D} D$, $\rho {\psi}' \to \bar{D}^* D$ or
$\bar D D^*$, $\rho {\psi}' \to \bar{D}^* D^*$, $\rho \chi_{c} \to \bar{D} D$, 
$\rho \chi_{c} \to \bar{D}^* D$ or $\bar D D^*$, and $\rho \chi_{c} \to 
\bar{D}^* D^*$. Here $D$ stands for $D^+$ or $D^0$, $\bar{D}$ for $D^-$ or 
$\bar{D}^0$, $D^*$ for $D^{^*+}$ or $D^{*0}$, and $\bar{D}^*$ for $D^{*-}$ or 
$\bar{D}^{*0}$. For the reaction that produces $\bar D^* D$ or $\bar D D^*$, 
the cross section for the production of $\bar D D^*$ equals the one of $\bar 
D^* D$. No matter what the temperature is, the reactions,  $\pi J/\psi \to 
\bar{D}^{*} D$, $\pi J/\psi \to\bar{D}^{*} D^{*}$, $\pi {\psi}' \to \bar{D}^* 
D^*$, $\pi \chi_{c} \to \bar{D}^* D^*$, $\rho J/\psi \to \bar{D}^* D$, and 
$\rho J/\psi \to \bar{D}^* D^*$, are endothermic. No reactions are all 
exothermic from $T=0.6T_{\rm c}$ to $T_{\rm c}$. To display in-medium 
modification of dissociation, we plot the cross sections at $T=0$ for reference
even though they differ from the results obtained in Ref. \cite{BSWX} from the 
potential that includes the color Coulomb, spin-spin hyperfine and linear 
confinement terms.

\vspace{0.5cm}
\leftline {\sl 4.2. Endothermic reactions of $\pi {\psi}'$, $\pi \chi_c$, $\rho
{\psi}'$ and $\rho \chi_c$}
\vspace{0.5cm}

The cross sections for endothermic $\pi {\psi}'$ and $\pi \chi_c$ reactions in 
Figs. 4-7 show that peak cross sections increase from $T/T_{\rm c}=0$ to 0.75 
and decrease from  $T/T_{\rm c}=0.75$ to 0.95 except the increasing peak cross 
section from  $T/T_{\rm c}=0.9$ to 0.95 in Fig. 5 for $\pi {\psi}' \to \bar D^*
D^*$. While temperature increases from zero, the large-distance potential given
by the first term in Eq. (1) becomes smaller and smaller, and the Schr\"odinger
equation with the potential shown by Eq. (1) produces increasing meson radius. 
The increase of initial-meson radii leads to peak cross sections increasing 
from  $T/T_{\rm c}=0$ to 0.75. With increasing temperature, it becomes more 
difficult to combine the interchanged quarks and the antiquarks to form mesons 
$q \bar c$ and $c \bar q$ due to weakening confinement. This leads to peak 
cross sections decreasing from  $T/T_{\rm c}=0.75$ to 0.95. This also explains 
the decrease of the peak cross section from $T/T_{\rm c}=0.85$ to 0.9 in Fig. 
13 for $\rho {\psi}' \to \bar D^* D^*$, from $T/T_{\rm c}=0.9$ to 0.95 in Fig. 
11 for $\rho {\psi}' \to \bar D D$, Fig. 12 for $\rho {\psi}' \to \bar D^* D$, 
Fig. 14 for $\rho \chi_c \to \bar D D$ and Fig. 15 for $\rho \chi_c \to \bar 
D^* D$, or from $T/T_{\rm c}=0.85$ to 0.95 in Fig. 16 for $\rho \chi_c \to \bar
D^* D^*$.

We have already known that ${\psi}'$ and $\chi_c$ have very similar temperature
dependence in mass. The $\pi {\psi}'$ ($\rho {\psi}'$) and $\pi \chi_c$ ($\rho 
\chi_c$) reactions are almost identical in threshold energy, but are different 
in cross section even though the final mesons in the $\pi {\psi}'$ ($\rho 
{\psi}'$) reactions are the same as those in the $\pi \chi_c$ ($\rho \chi_c$) 
reactions. The difference is clearly attributed to the different quantum 
numbers of ${\psi}'$ and $\chi_c$. Particularly, when $T/T_{\rm c}$ goes from 
0.9 to 0.95, the increase of the peak cross section of $\pi {\psi}' \to \bar 
D^* D^*$ in Fig. 5 ($\rho {\psi}' \to \bar D^* D^*$ in Fig. 13) is against the 
decrease of the peak cross section of $\pi \chi_c \to \bar D^* D^*$ in Fig. 7 
($\rho \chi_c \to \bar D^* D^*$ in Fig. 16). This relates to the node in the 
${\psi}'$ wave function. The node leads to cancellation between the negative 
wave function on the left of the node and the positive wave function on the 
right of the node in the integration involved in the transition amplitude. 
While the cancellation at $T=0.95T_{\rm c}$ is less than at $T=0.9T_{\rm c}$, 
the peak cross sections of $\pi {\psi}' \to \bar{D}^* D^*$ and $\rho {\psi}' 
\to \bar{D}^* D^*$ rise from $T/T_{\rm c}=$ 0.9 to 0.95.

\vspace{0.5cm}
\leftline {\sl 4.3. Exothermic reactions of $\pi {\psi}'$, $\pi \chi_c$, $\rho 
{\psi}'$, $\rho \chi_c$ and $\rho J/\psi$}
\vspace{0.5cm}

Since cross sections for exothermic reactions at threshold energies are 
infinite, we start calculating the cross sections at $\sqrt s = m_{q \bar q} 
+m_{c \bar c}+ \triangle \sqrt s $ with $\triangle \sqrt s = {10}^{-4}$ GeV. At
the energies the cross sections correspond to the curve tops, for example, the 
top of the curve for $T/T_{\rm c}=0.65$ displayed in Fig. 4, and from Eq. (12) 
we have
\begin{equation}
\vec {P^2} \approx 2\bigtriangleup \sqrt{s}/{(\frac{1}{m_{q \bar q}}
+\frac{1}{m_{c \bar c}})},
\end{equation}
which indicates that $\vec{P}^2$ decreases with decreasing meson mass. In 
Section 2 we have obtained that the meson masses decrease with increasing 
temperature. Hence, $|\vec P|$ decreases while temperature increases. At 
$T=0.6T_{\rm c}$, $|\vec P|$ is estimated at about ${10}^{-2}$ GeV. The 
influence of $\vec P$ on the relative momenta $\vec p_{ab}$ is thus negligible.
Finally, $\vec p_{ab}$ is governed by $\vec {P}^{\prime}$ that obeys
\begin{equation}
m_{q \bar q}+m_{c \bar c} + \Delta \sqrt{s} = \sqrt{{m^2_{q \bar c}}
+{{\vec {P'}}^2}}+\sqrt{{m^2_{c \bar q}}+{{\vec {P'}}^2}}.
\end{equation}
Since $\Delta \sqrt {s}$ is very small, we have
\begin{equation}
{\vec {P'}}^2 \approx 2(m_{q \bar q} + m_{c \bar c} - m_{q \bar c} - 
m_{c \bar q})/(\frac{1}{m_{q \bar c}} + \frac{1}{m_{c \bar q}}).
\end{equation}
With increasing temperature the value of $m_{q \bar q} + m_{c \bar c} - 
m_{q \bar c} - m_{c \bar q}$ decreases in the region where reactions are 
exothermic, and becomes zero at a temperature; meanwhile, $|\vec {P'}|$ 
decreases so that the absolute value of the relative-motion part of 
${\psi}_{ab} (\vec p_{ab})$ gets larger and larger, and the factor $\sqrt 
{E_{q \bar q}E_{c \bar c}E_{q \bar c}E_{c \bar q}}$ in the transition amplitude
decreases. Simple calculations show that the factor $\frac{1}{s}\frac{|\vec 
{P}^{\prime}|}{|\vec {P}|}$ generally decreases with increasing temperature in 
the region where reactions are exothermic. The product of the increasing 
absolute value of the relative-motion part of
${\psi}_{ab}(\vec p_{ab})$ and the decreasing $\frac{1}{s}\frac{|\vec 
{P}^{\prime}|}{|\vec {P}|}{E_{q \bar q}E_{c \bar c}E_{q \bar c}E_{c \bar q}}$ 
leads to that the cross section at $\sqrt s = m_{q \bar q}+m_{c \bar c}  + 
\Delta \sqrt s$ first increases and then decreases as shown in Figs. 6, 8, and 
16, or first decreases and then increases in Figs. 4 and 11 through 15.

\vspace{0.5cm}
\leftline {\sl 4.4. Endothermic reactions of $\pi J/\psi$ and $\rho J/\psi$}
\vspace{0.5cm}

As shown in Figs. 2-3 and 8-10, the peak cross sections of the endothermic $\pi
J/\psi$ and $\rho J/\psi$ reactions decrease from $T=0$ to $0.85 T_{\rm c}$ and
increase from $T=0.85 T_{\rm c}$ to $0.95 T_{\rm c}$. The behavior of the peak 
cross sections is different from that of the endothermic reaction $\pi \chi_c 
\to \bar D^* D^*$, which is an increase from $T=0$ to $0.75 T_{\rm c}$ and a 
decrease from $T=0.75 T_{\rm c}$ to $0.95 T_{\rm c}$. The behavior must come 
from a special point, and as demonstrated below, the special point is just the 
temperature dependence of the mass and radius of $J/\psi$.

First, we discuss cross sections for $\pi J/\psi$ and $\rho J/\psi$ 
dissociation in the region $0 \leq T/T_{\rm c} \leq 0.85$. In Section 2 we have
obtained the slow decrease of ${\psi}'$, $\chi_c$ and $D^*$ masses and the very
slow decrease of $J/\psi$ and $D$ masses when $T/T_{\rm c}$ increases from 0 to
0.85. It is shown in Fig. 2 of Ref. \cite{ZXG} that the $\pi$ mass decreases 
slowly for $0.6T_{\rm c} \leq T < 0.78T_{\rm c}$ and the $\rho$ mass does 
rapidly. Then, $\frac{1}{s}\frac{|\vec {P'}|}{|\vec P|}$ keeps roughly 
unchanged or decreases. Since the $J/\psi$ radius changes very slowly in 
comparison to the increase of the $\pi$ or $\rho$ radius, only one increasing 
initial-meson radius causes the increase of $|{\cal M}_{\rm fi}|^2$, and the 
increase can not overcome the amount reduced by the weakening confinement with 
increasing temperature, i.e. $|{\cal M}_{\rm fi}|^2$ decreases. Finally, the 
product of $\frac{1}{s}\frac{|\vec {P'}|}{|\vec P|}$ and 
$|{\cal M}_{\rm fi}|^2$ leads to the decrease of the peak cross sections of the
endothermic $\pi J/\psi$ and $\rho J/\psi$ reactions when $T/T_{\rm c}$ 
increases from 0 to 0.85.

Next, we discuss cross sections for $\pi J/\psi$ and $\rho J/\psi$ dissociation
in the region $0.85 \leq T/T_{\rm c} < 1$. The factor $\frac{1}{s}\frac{|\vec 
{P'}|}{|\vec P|}$ increases by about $50 \%$ from $T=0.85T_{\rm c}$ to 
$0.9T_{\rm c}$ and by about $100 \%$ from $T=0.9T_{\rm c}$ to $0.95T_{\rm c}$. 
When temperature increases from $T=0.85T_{\rm c}$ to $0.95T_{\rm c}$, not only 
the $\pi$ and $\rho$ radii increase, but also the $J/\psi$ radius apparently 
does. Now the increase of $|{\cal M}_{\rm fi}|^2$ caused by the increasing 
radii of the two initial mesons overcomes the amount reduced by the weakening 
confinement with increasing temperature. Finally, the peak cross sections of 
the endothermic $\pi J/\psi$ and $\rho J/\psi$ reactions increase from 
$T/T_{\rm c}=0.85$ to 0.95.

\vspace{0.5cm}
\leftline {\sl 4.5. Comparison of $\pi$+ charmonium reactions and $\rho$+ 
charmonium reactions}
\vspace{0.5cm}

We make a comparison of peak cross sections of endothermic $\pi$ + charmonium 
reactions with ones of endothermic $\rho$ + charmonium reactions. We find that 
the peak cross section of $\pi J/\psi \to \bar{D}^* D$ ($\pi \chi_c \to \bar 
D^* D$) is larger than that of $\rho J/\psi \to \bar{D}^* D$ ($\rho \chi_c \to 
\bar D^* D$) at the same temperature while the peak cross section of $\pi$ + 
charmonium $\to \bar D^* + D^*$ ($\pi {\psi}' \to \bar D^* D$) is smaller than 
the one of $\rho$ + charmonium $\to \bar D^* + D^*$ ($\rho {\psi}' \to \bar D^*
D$). Exothermic $\pi$ + charmonium reactions and exothermic $\rho$+ charmonium 
reactions are compared about their cross sections at $\sqrt s = m_{q \bar q} + 
m_{c \bar c} + 10^{-4}$ GeV (the largest cross section shown in each curve). It
is shown in Figs. 4, 6, 12 and 15 that the cross section of $\pi +$ charmonium 
$\to \bar D^* + D$ at $T/T_{\rm c}=0.65$ or 0.75 is larger than the one of 
$\rho +$ charmonium $\to \bar D^* + D$ at the same temperature. It is 
meaningless to compare endothermic $\pi$ + charmonium reactions (e.g. $\pi 
{\psi}' \to \bar D^* D$ at $T/T_{\rm c}=0.85$) with exothermic $\rho$ + 
charmonium reactions (e.g. $\rho {\psi}' \to \bar D^* D$ at the same 
temperature).

Regarding endothermic reactions we examine $\vec P$ and $\vec P'$ at $\sqrt s =
m_{q \bar c} + m_{c \bar q} + d_0$ at which the peak cross sections are given 
and $d_0 \ll m_{q \bar c} + m_{c \bar q}$. Eq. (13) yields
\begin{equation}
\vec {P'}^2 \approx 2d_0/(\frac{1}{m_{q \bar c}} + \frac{1}{m_{c \bar q}}).
\end{equation}
From an endothermic $\pi$ + charmonium reaction to an endothermic $\rho$ + 
charmonium reaction at the same temperature, $m_{q \bar c}$ and $m_{c \bar q}$ 
do not change, and $d_0$ decreases or does not change. Accordingly, 
$\vec {P'}^2$ is reduced or unchanged. $\sqrt s$ is related to initial-meson 
energies by
\begin{equation}
m_{q \bar c} + m_{c \bar q} + d_0 = \sqrt {m_{q \bar q}^2 + {\vec P}^2} + 
\sqrt {m_{c \bar c}^2 + {\vec P}^2},
\end{equation}
which let ${\vec P}^2$ decrease from the endothermic $\pi$ + charmonium 
reaction to the endothermic $\rho$ + charmonium reaction. As a result the
absolute value of the relative-motion part of ${\psi}_{ab}({\vec p}_{ab})$ in 
the transition amplitude increases.

In the transition amplitude $\psi_{ab}$ is the product of color, spin, flavor 
and momentum-space wave functions of the relative motion of constituents $a$ 
and $b$. The transition amplitude involves the spin factor: the matrix elements
of ${\vec s_a} \cdot {\vec s_b}$ of the spin-spin interaction and the overlap 
of the spin wave function of final mesons and the one of initial mesons. The 
difference of the transition amplitudes for $\pi$ + charmonium $\to \bar D^* + 
D $ and for $\rho$ + charmonium $\to \bar D^* + D$ is caused by not only the 
difference of the $\pi$ and $\rho$ masses (the subsequent changes of $\sqrt 
{E_{q \bar q}E_{c \bar c}E_{q \bar c}E_{c \bar q}}$, ${\psi}_{ab}
(\vec p_{ab})$, and $|{\vec P}'|/s|\vec P|$) but also the difference of the 
spin factors for the two reactions. Particularly, $\pi$ and $\rho$ become 
almost degenerate in mass at $T/T_{\rm c} \geq 0.9$. Then, only the spin 
factors made the peak cross section of $\pi$ + charmonium $\to \bar D^* + D$ 
larger or smaller than the one of $\rho$ + charmonium $\to \bar D^* + D$. In 
case the final mesons are $\bar{D}^*$ and $D^*$, the total spin of $\pi$ and 
charmonium is 1 that differs from the total spins 0, 1 and 2 of $\rho$ and 
charmonium. The spin factors plus the additional contribution of the channels 
of the total spins 0 and 2 of $\rho$ and charmonium make the peak cross section
of $\pi$ + charmonium $\to$ $\bar{D}^*$ + $D^*$ smaller than the one of $\rho$ 
+ charmonium $\to$ $\bar{D}^*$ + $D^*$ at $T/T_{\rm c}=0.9$ and 0.95. The 
difference between the cross sections at $\sqrt s = m_{q \bar q} + m_{c \bar c}
+10^{-4}$ GeV for exothermic $\pi$ + charmonium $\to \bar D^* +D$ and for 
exothermic $\rho$ + charmonium $\to \bar D^* +D$ can be understood similarly 
with Eqs. (18) and (19).

\vspace{0.5cm}
\leftline{\bf 5. Results pertinent to smearing in the spin-spin interaction}
\vspace{0.5cm}

We have solved the Schr\"odinger equation with the central spin-independent
potential $V_{\rm si}$ in Section 2. If we use the spin-spin interaction given
in Eq. (3) in the Schr\"odinger equation, the delta function in the first 
term of the $V_{\rm ss}$ expression cannot be correctly dealt with. However, 
in the present section
we take smearing in the spin-spin interaction by the regularization
$\delta^3 (\vec {r}) \to \frac {d^3}{\pi^{3/2}}\exp (-d^2r^2)$, and the smeared
spin-spin interaction
\begin{equation}
V_{\rm ss}=
- \frac {\vec {\lambda}_a}{2} \cdot \frac {\vec {\lambda}_b}{2}
\frac {16\pi^2}{25} \frac {d^3}{\pi^{3/2}}\exp (-d^2r^2)
\frac {\vec {s}_a \cdot \vec {s}_b}{m_am_b}
+ \frac {\vec {\lambda}_a}{2} \cdot \frac {\vec {\lambda}_b}{2}
  \frac {4\pi}{25} \frac {1}{r}
\frac {d^2v(\lambda r)}{dr^2} \frac {\vec {s}_a \cdot \vec {s}_b}{m_am_b} ,
\end{equation}
can be used in the Schr\"odinger equation. Such smearing actually includes 
relativistic effects \cite{WSB01,BS92,Wong}. The quantity $d$ is related to
quark masses by
\begin{displaymath}
d^2=\sigma^2_0 \left[ \frac {1}{2} + \frac {1}{2} \left( 
\frac {4m_am_b}{(m_a+m_b)^2} \right)^4 \right]
+\sigma^2_1 \left( \frac {2m_am_b}{m_a+m_b} \right)^2 ,
\end{displaymath}
where $\sigma_0=0.15$ GeV and $\sigma_1=0.705$. By solving the Schr\"odinger 
equation with the central spin-independent potential and the smeared spin-spin
interaction, we obtain meson masses and quark-antiquark relative-motion wave 
functions that differ by different mesons. The $\rho$ wave function is near 
that obtained by the Schr\"odinger equation with only $V_{\rm si}$, but
the $\pi$ wave function is not. Since the smeared spin-spin interaction is 
used, the pion becomes a tight bound state of a quark and an antiquark. Then,
the $\pi$ radius increases slowly from 0.6 to 0.9 of $T/T_c$ and quickly from
0.9 to 1 of $T/T_c$.

The mass splittings at $T=0$ are
$m_\rho - m_\pi = 0.6294$ GeV and $m_{K^\ast} - m_K = 0.39865$ GeV which
are closer to the experimental data 0.6304 GeV and 0.3963 GeV than the 
results (0.5989 GeV and 0.3833 GeV) \cite{ZXG}
of the Schr\"odinger equation with only the
central spin-independent potential. At T=0 we also get 3.13509 GeV, 3.69248 
GeV, 3.50578 GeV, 1.90578 GeV, 2.05274 GeV, 1.9614 GeV, and 2.13804 GeV as the
masses of $J/\psi$, $\psi^\prime$, $\chi_c$, $D$, $D^\ast$, $D_s$, and $D^*_s$,
respectively, compared to the measured values 3.096916 GeV, 3.68609 GeV,
3.5253 GeV, 1.86722 GeV, 2.00861 GeV, 1.96847 GeV, and 2.1123 GeV 
\cite{KN2010}. For $0.6 \leq T/T_c < 1$ temperature dependence of the $\rho$,
$K^\ast$, $J/\psi$, $\psi^\prime$, $\chi_c$, $D$, $D^\ast$, $D_s$ and $D^*_s$
masses obtained with the potential
$V_{\rm si}+V_{\rm ss}$ is very close to that obtained with only $V_{\rm si}$.
Therefore, the parametrizations given in Eqs. (4)-(10) are also valid in the
present section, and the $\rho$ and $K^\ast$ masses in units of GeV in the
region $0.6 \leq T/T_c < 1$ are parametrized as
\begin{equation}
m_{\rho}=0.73\left[ 1-\left( \frac{T}{0.992T_c} \right)^{3.67} \right]^{0.989},
\end{equation}
\begin{equation}
m_{K^\ast}=0.84 \left[ 1-\left( \frac{T}{1.05T_c} \right)^{4.16} \right] .
\end{equation}
However, the $\pi$, $K$ and $\eta$ masses obtained with 
$V_{\rm si}+V_{\rm ss}$ are smaller than those obtained
with only $V_{\rm si}$. They are plotted in Fig. 17 for $0.6 \leq T/T_c <1$
and are parametrized as
\begin{equation}
m_{\pi}=0.24\left[ 1-\left( \frac{T}{0.97T_c} \right)^{3.81} \right]^{0.51},
\end{equation}
\begin{equation}
m_K=0.46 \left[ 1-\left( \frac{T}{1.04T_c} \right)^{8.58} \right]^{0.88},
\end{equation}
\begin{equation}
m_{\eta}=0.55\left[ 1-\left( \frac{T}{1.01T_c} \right)^{3.11} \right]^{0.29},
\end{equation}
in units of GeV.

Corresponding to the $\pi$ mass, the reactions $\pi J/\psi \to \bar {D}^* D$,
$\pi J/\psi \to \bar {D}^* D^*$, $\pi \psi^\prime \to \bar {D}^* D$,
$\pi \psi^\prime \to \bar {D}^* D^*$, $\pi \chi_c \to \bar {D}^* D$, and
$\pi \chi_c \to \bar {D}^* D^*$ are all endothermic. Cross sections for the
reactions are calculated with the experimental meson masses at $T=0$,
the temperature-dependent meson masses for $0.6 \leq T/T_c <1$, the 
quark-antiquark relative-motion wave functions and the Fourier transform of the
sum of the central spin-independent potential and the smeared spin-spin
interaction
\begin{eqnarray}
V_{ab}\left( \vec {Q}\right) &=& -\frac{ \vec {\lambda }_{a}}{2}
\cdot \frac{\vec {\lambda }_{b}}{2}\frac{3}{4} D
\left[ 1.3- \left( \frac {T}{T_{\rm c}} \right)^4 \right]
\left[ (2\pi)^3\delta^3 (\vec {Q}) - \frac {8\pi}{Q}
\int^\infty_0 dr \frac {r\sin (Qr)}{\exp (2Ar)+1} \right]
                               \notag \\
& &
+\frac{ \vec {\lambda }_{a}}{2} \cdot \frac{\vec {\lambda }_{b}}{2} 64 \pi E
\int^\infty_0 dq \frac {\rho (q^2) -\frac {K}{q^2}}{(E^2+Q^2+q^2)^2-4Q^2q^2}
                               \notag \\
& & -\frac{\vec {\lambda }_{a}}{2}
\cdot \frac{\vec {\lambda }_{b}}{2}\frac{16\pi ^{2}}{25} 
\exp \left( -\frac {Q^2}{4d^2} \right)
\frac{\vec {s}_{a}\cdot \vec {s}_{b}}{m_{a}m_{b}}
                               \notag \\
& &
+\frac{\vec {\lambda }_{a}}{2}\cdot \frac{\vec {\lambda }_{b}}{
2}\frac{16\pi ^{2}\lambda }{25Q}\int_{0}^{\infty}dx\frac{d^{2}v\left(
x\right) }{dx^{2}}\sin \left( \frac{Q}{\lambda }x\right)
\frac{\vec {s}_{a}\cdot \vec {s}_{b}}{m_{a}m_{b}} .
                                 \notag \\
\end{eqnarray}
We plot the cross sections for the
reactions in Figs. 18-23. The cross sections differ from those shown in Figs. 
2-7. For example, the reactions $\pi \psi^\prime \to \bar {D}^* D$ and
$\pi \chi_c \to \bar {D}^* D$ at $T/T_c=0.65$ and 0.75 are exothermic in Figs.
4 and 6 but endothermic in Figs. 20 and 22. One feature for each reaction is
that the peak cross section at $T/T_c=0.95$ is always larger than the ones
at the other five temperatures $T/T_c=0$, 0.65, 0.75, 0.85, 0.9. This is
caused by the $\pi$ and $J/\psi$ radii which increase faster from 0.9 to 1 of
$T/T_c$ than from 0.6 to 0.9 of $T/T_c$.
The numerical cross sections for endothermic reactions are parametrized as
\begin{eqnarray}
\sigma^{\rm unpol}(\sqrt {s},T) & = &
a_1 \left( \frac {\sqrt {s} -\sqrt {s_0}}{b_1} \right)^{c_1}
\exp \left[ c_1 \left( 1-\frac {\sqrt {s} -\sqrt {s_0}}{b_1} \right) \right]
                   \notag   \\
& & + a_2 \left( \frac {\sqrt {s} -\sqrt {s_0}}{b_2} \right)^{c_2}
\exp \left[ c_2 \left( 1-\frac {\sqrt {s} -\sqrt {s_0}}{b_2} \right) \right] ,
\end{eqnarray}
where $\sqrt {s_0}$ is the threshold energy, and $a_1$, $b_1$, $c_1$, $a_2$, 
$b_2$ and $c_2$ are parameters. 
Determination of parameter values needs time-consuming computations, and 
the values are listed in Tables 1-3.

Since the $\rho$, $J/\psi$, $\psi^\prime$, $\chi_c$, $D$ and $D^*$
masses determined by the Schr\"odinger equation in the case of
$V_{\rm si}+V_{\rm ss}$ are very close to the masses determined by the 
Schr\"odinger equation in the case of only $V_{\rm si}$, a $\rho$-charmonium
dissociation reaction which is endothermic (exothermic) in the former case 
is also endothermic
(exothermic) in the latter case, and $\frac {1}{s}\frac {\mid
\vec {P}^\prime \mid}{\mid \vec {P} \mid}$ in the two cases are quite close.
Since $\sqrt s$ dependence of the cross section for a quark-interchange-induced
reaction is mainly determined by 
$\frac {1}{s}\frac {\mid \vec {P}^\prime \mid}{\mid \vec {P} \mid}$ \cite{LX},
$\sqrt s$ dependence of the cross section for
a $\rho$-charmonium reaction in the former case is
similar to one in the latter case. Hence, we do not plot $\rho$-charmonium 
dissociation cross sections in the case of $V_{\rm si}+V_{\rm ss}$, 
but present parametrizations of these cross 
sections. The cross section for the exothermic reaction 
$q \bar q (S_{q \bar q};\vec P)+c \bar c (S_{c \bar c};-\vec P)\to q \bar 
c(S_{q \bar c};\vec P')+c \bar q(S_{c \bar q};-\vec P')$ can be related to the 
endothermic reaction $q \bar c + c \bar q \to q \bar q + c \bar c$ by the 
detailed balance
\begin{eqnarray}
\sigma^{\rm unpol}_{q \bar q + c \bar c \to q \bar c + c \bar q}= 
\frac{(2S_{q \bar c} +1)(2S_{c \bar q} +1)}{(2S_{q \bar q} +1)
(2S_{c \bar c} +1)}\frac{{\vec {P'}}^2}{{\vec P}^2}
\sigma^{\rm unpol}_{q \bar c + c \bar q \to q \bar q + c \bar c} ,
\end{eqnarray}
where $S_{q \bar c}$ and $S_{c \bar q}$ are the spins of $q \bar c$ and $c \bar
q$, respectively. It is then correct to choose the following parametrization 
for the exothermic reaction
\begin{eqnarray}
\sigma^{\rm unpol}(\sqrt {s},T) & = &
\frac{{\vec {P'}}^2}{{\vec P}^2}
\left\{a_1 \left( \frac {\sqrt {s} -\sqrt {s_0}}{b_1} \right)^{c_1}
\exp \left[ c_1 \left( 1-\frac {\sqrt {s} -\sqrt {s_0}}{b_1} 
\right) \right]\right.
                   \notag   \\
& & \left. + a_2 \left( \frac {\sqrt {s} -\sqrt {s_0}}{b_2} \right)^{c_2}
\exp \left[ c_2 \left( 1-\frac {\sqrt {s} -\sqrt {s_0}}{b_2} 
\right) \right]\right\} .
\end{eqnarray}
Parameter values are listed in Tables 4-6.

\vspace{0.5cm}
\leftline{\bf 6. Procedure}
\vspace{0.5cm}

The curves shown in Figs. 2-16 and 18-23
correspond to the zero temperature and  the five
nonzero temperatures $T_1=0.65T_{\rm c}$, $T_2=0.75T_{\rm c}$, 
$T_3=0.85T_{\rm c}$, $T_4=0.9T_{\rm c}$ and  $T_5=0.95T_{\rm c}$.
We now present a procedure on how to obtain the unpolarized cross section 
for $q\bar {q} + c\bar {c} \to q\bar {c} + c\bar {q}$
for $0.65T_{\rm c} \leq T < T_{\rm c}$ from the curves.

First, we state the procedure while a reaction at $T_i$ and $T_{i+1}$ is 
endothermic. We denote by $\sqrt {s_{\rm p}}$ the square root of the Mandelstam
variable corresponding to the peak cross section. $d_0 = \sqrt {s_{\rm p}} - 
\sqrt {s_0}$ is the difference of $\sqrt {s_{\rm p}}$ with respect to the 
threshold energy $\sqrt {s_0}$. Let $\sqrt {s_{\rm z}}$ be the square root of 
the Mandelstam variable at which the cross section is 1/100 of the peak cross 
section, and $\sqrt {s_{\rm z}}>\sqrt {s_{\rm p}}>\sqrt {s_0}$. 
$d_0$, $\sqrt {s_{\rm z}}$, $\sqrt {s_0}$, and $\sqrt {s_{\rm p}}$ at $T_i$ are
indicated by $d_{0i}$, $\sqrt {s_{{\rm z}i}}$, $\sqrt {s_{0i}}$, and $\sqrt 
{s_{{\rm p}i}}$, respectively.  $d_{0i}$ and $\sqrt{s_{{\rm z}i}}$ can be found
in Tables 1-6. $\sqrt {s_{0i}}$ can be obtained from the mass parametrizations 
in Section 2, and $\sqrt {s_{{\rm p}i}}= \sqrt {s_{0i}} + d_{0i}$.
For $T_i\leq T \leq T_{i+1}$ ($i$=1, 2, 3 or 4) we take the linear 
interpolation between the two peaks at $T_i$ and $T_{i+1}$ to estimate $d_0$ 
and $\sqrt{s_{\rm z}}$ of the cross section at $T$,
\begin{equation}
d_0=\frac{d_{0i+1}-d_{0i}}{T_{i+1}-T_i}(T-T_i)+d_{0i},
\end{equation}
\begin{equation}
\sqrt{s_{\rm z}}=
\frac{\sqrt{s_{{\rm z}i+1}}-\sqrt{s_{{\rm z}i}}}{T_{i+1}-T_i}(T-T_i)
+\sqrt{s_{{\rm z}i}}.
\end{equation}
$\sqrt{s_0}$ is the sum of final-meson masses, and parametrizations of the 
masses have been given by Eqs. (7) and (8).
The square root of the Mandelstam variable corresponding to the peak cross 
section is $\sqrt{s_{\rm p}}=\sqrt{s_0}+d_0$. We define a ratio
\begin{equation}
\zeta=
\left\{ \begin{array}{c}
\frac{\sqrt{s}-\sqrt{s_{\rm p}}}{\sqrt{s_0}-\sqrt{s_{\rm p}}} ~~~~ {\rm if}~
\sqrt{s_0}\leq\sqrt{s}\leq\sqrt{s_{\rm p}}
\\
\frac{\sqrt{s}-\sqrt{s_{\rm p}}}{\sqrt{s_{\rm z}}-\sqrt{s_{\rm p}}} ~~~~
{\rm if}~ \sqrt{s_{\rm p}}<\sqrt{s}\leq\sqrt{s_{\rm z}}.
\end{array} \right.
\end{equation}
The ratio $\zeta$ is on the closed interval [0,1], and corresponds to $\sqrt s$
and $T$. In the cross section curve at $T_i$, we can find a point 
$(\sqrt {s_i}, \sigma^{\rm unpol}_i(\sqrt {s_i}, T_i))$ which gives the same 
ratio,
\begin{equation}
\zeta=
\left\{ \begin{array}{c}
\frac{\sqrt{s_i}-\sqrt{s_{{\rm p}i}}}{\sqrt{s_{0i}}-\sqrt{s_{{\rm p}i}}} ~~~~ 
{\rm if}~
\sqrt{s_{0i}}\leq\sqrt{s_i}\leq\sqrt{s_{{\rm p}i}}
\\
\frac{\sqrt{s_i}-\sqrt{s_{{\rm p}i}}}{\sqrt{s_{{\rm z}i}}-\sqrt{s_{{\rm p}i}}} 
~~~~ {\rm if}~ \sqrt{s_{{\rm p}i}}<\sqrt{s_i}\leq\sqrt{s_{{\rm z}i}},
\end{array} \right.
\end{equation}
and in the cross section curve at $T_{i+1}$, we find a point $(\sqrt {s_{i+1}},
\sigma^{\rm unpol}_{i+1}(\sqrt {s_{i+1}}, T_{i+1}))$ which also gives the ratio
\begin{equation}
\zeta=
\left\{ \begin{array}{c}
\frac{\sqrt{s_{i+1}}-\sqrt{s_{{\rm p}{i+1}}}}{\sqrt{s_{0{i+1}}}
-\sqrt{s_{{\rm p}{i+1}}}} ~~~~ {\rm if}~
\sqrt{s_{0{i+1}}}\leq\sqrt{s_{i+1}}\leq\sqrt{s_{{\rm p}{i+1}}}
\\
\frac{\sqrt{s_{i+1}}-\sqrt{s_{{\rm p}{i+1}}}}{\sqrt{s_{{\rm z}{i+1}}}
-\sqrt{s_{{\rm p}{i+1}}}} ~~~~
{\rm if}~ \sqrt{s_{{\rm p}{i+1}}}<\sqrt{s_{i+1}}\leq\sqrt{s_{{\rm z}{i+1}}}.
\end{array} \right.
\end{equation}
The regions $\sqrt {s_{0i}} \leq \sqrt {s_i} \leq \sqrt {s_{{\rm p}i}}$ and 
$\sqrt {s_{0i+1}} \leq \sqrt {s_{i+1}} \leq \sqrt {s_{{\rm p}{i+1}}}$ 
($\sqrt {s_{{\rm p}i}} < \sqrt {s_i} \leq \sqrt {s_{{\rm z}i}}$ and 
$\sqrt {s_{{\rm p}i+1}} < \sqrt {s_{i+1}} \leq \sqrt {s_{{\rm z}{i+1}}}$) 
correspond to $\sqrt {s_0} \leq \sqrt s \leq \sqrt {s_{\rm p}}$ 
($\sqrt {s_{\rm p}} < \sqrt s \leq \sqrt {s_{\rm z}}$). In terms of the ratio
$\sqrt {s_i}$ and $\sqrt {s_{i+1}}$ are expressed as
\begin{equation}
\sqrt {s_i}=
\left\{ \begin{array}{c}
\sqrt {s_{{\rm p}i}} + \zeta(\sqrt {s_{0i}}- \sqrt {s_{{\rm p}i}}) ~~~~ 
{\rm if}~ \sqrt{s_0}\leq\sqrt{s}\leq\sqrt{s_{\rm p}}
\\
\sqrt {s_{{\rm p}i}} + \zeta(\sqrt {s_{{\rm z}i}}- \sqrt {s_{{\rm p}i}}) ~~~~ 
{\rm if}~ \sqrt{s_{\rm p}}<\sqrt{s}\leq\sqrt{s_{\rm z}} ,
\end{array} \right. 
\end{equation}
\begin{equation}
\sqrt {s_{i+1}}=
\left\{ \begin{array}{c}
\sqrt {s_{{\rm p}{i+1}}} + \zeta(\sqrt {s_{0{i+1}}}- \sqrt {s_{{\rm p}{i+1}}}) 
~~~~ {\rm if}~ \sqrt{s_0}\leq\sqrt{s}\leq\sqrt{s_{\rm p}}
\\
\sqrt {s_{{\rm p}{i+1}}} + \zeta(\sqrt {s_{{\rm z}{i+1}}}- \sqrt {s_{{\rm p}
{i+1}}}) ~~~~ {\rm if}~ \sqrt{s_{\rm p}}<\sqrt{s}\leq\sqrt{s_{\rm z}} .
\end{array} \right. 
\end{equation}
The linear interpolation between the two points provides the unpolarized cross 
section at $\sqrt s$ for $T_i \leq T\leq T_{i+1}$,
\begin{displaymath}
\sigma ^{\rm unpol} (\sqrt {s},T)=
\end{displaymath}
\begin{equation}
\left\{ \begin{array}{c}
\frac{\sigma^{\rm unpol}_{i+1}(\sqrt {s_{i+1}}, T_{i+1})
-\sigma^{\rm unpol}_i(\sqrt {s_i}, T_i)}{T_{i+1}-T_i}(T-T_i)
+\sigma^{\rm unpol}_i(\sqrt {s_i}, T_i) ~~~~ {\rm if}~
\sqrt{s_0}\leq\sqrt{s}\leq\sqrt{s_{\rm z}}
\\
0~~~~~~~~~~~~~~~~~~~~~~~~~~~~~~~~~~~~~~~~~~~~~~~~~~~~~~~~~~~~~~ 
{\rm if}~ \sqrt{s}>\sqrt{s_{\rm z}}.
\end{array} \right.
\end{equation}
where $\sigma ^{\rm unpol}_i$ and $\sigma ^{\rm unpol}_{i+1}$ are shown in
Figs. 2-16 and 18-23, and are given by Eq. (30) together with the 
parameters listed in
Tables 1-6. For $0.95T_{\rm c}<T < T_{\rm c}$ Eqs. (33)-(40) still apply to
obtaining the unpolarized cross section so long as $T_i=T_4$ and $T_{i+1}=T_5$
are set.

Second, we state the procedure while a reaction at $T_i$ and $T_{i+1}$ is 
exothermic. Eqs. (33)-(40) are suited to endothermic reactions of which each 
has the zero
cross section at the threshold energy or at the infinite total energy of 
initial mesons and has one maximum cross section in its $\sqrt{s}$ dependence. 
The cross section for the exothermic reaction is infinite at the threshold 
energy. But
the quantity enclosed by the braces in Eq. (32) has the general $\sqrt{s}$
dependence of endothermic reactions.
Hence, Eqs. (33)-(39) apply to the quantity enclosed by the braces. $d_{0i}$ 
and $\sqrt {s_{{\rm z}i}}$ of the quantity are listed in Tables 4-6.
$\sqrt{s_0}$ equals the sum of initial-meson masses, and parametrizations of 
the masses have been given by Eqs. (4), (5), (6), and (24).  Eq. (40) now
gives the unpolarized cross section for the exothermic reaction at $T$ while 
$\sigma{^{\rm unpol}_i}$ and $\sigma{^{\rm unpol}_{i+1}}$ are the cross 
sections shown in Figs. 8 and 11-16, and are given by Eq. (32) together 
with the parameters listed in Tables 4-6. Since the infinity of the cross 
section at the threshold energy is intractable, do not let $\sqrt s$ equal 
$\sqrt {s_0}$ while fortran code is made. 

Third, we state the procedure while a reaction is exothermic at $T_i$ and 
endothermic at $T_{i+1}$. Eqs. (33)-(39) apply to the endothermic reaction at 
$T_{i+1}$ or the quantity enclosed by the braces in Eq. (32) at $T_i$. Relevant
$d_{0i}$, $\sqrt {s_{{\rm z}i}}$, $d_{0i+1}$, and $\sqrt {s_{{\rm z}i+1}}$ can 
be found in Tables 4-6. $\sqrt {s_{0i}}$ and $\sqrt {s_{0i+1}}$ equal the 
sum of initial-meson masses and the sum of final-meson masses, respectively. 
$\sqrt {s_0}$ at $T$ is the sum of final-meson masses for 
$m_{q \bar c} + m_{c \bar q} - m_{q \bar q} - m_{c \bar c} > 0$
or the sum of initial-meson masses for
$m_{q \bar c} + m_{c \bar q} - m_{q \bar q} - m_{c \bar c} < 0$.
After estimating $\sqrt {s_i}$ and $\sqrt {s_{i+1}}$, Eq. (40) is used to get 
the unpolarized cross section at $\sqrt s$ for $T_i\leq T \leq T_{i+1}$ while 
$\sigma ^{\rm unpol}_i(\sqrt {s_i}, T_i)$ is the cross section for the 
exothermic reaction and $\sigma ^{\rm unpol}_{i+1}(\sqrt {s_{i+1}}, T_{i+1})$ 
for the endothermic reaction.

\vspace{0.5cm}
\leftline{\bf 7. Summary}
\vspace{0.5cm}

The central spin-independent and temperature-dependent potential has been used 
in the Schr\"odinger equation to obtain the temperature-dependent masses of 
$J/\psi$, ${\psi}'$, $\chi_c$, $D$, $D^*$, $D_s$ and $D^*_s$ as well as the 
wave functions of quark-antiquark relative motion inside these mesons. The 
experimental masses of the mesons are reproduced by the potential at $T=0$. 
While temperature increases, the temperature dependence of the theoretical 
masses is: the ${\psi}'$ and $\chi_c$ masses decrease slowly, and the $J/\psi$ 
mass more slowly; each of the $D$ and $D_s$ masses keeps almost unchanged in a 
temperature region, and apparently falls off near $T_c$; the $D^*$ and $D^*_s$ 
masses first decrease slowly, and then apparently fall off. The prominent 
medium effects are: ${\psi}'$ and $\chi_c$ are degenerate in mass for 
$0.6T_{\rm c} < T < T_{\rm c}$; very near the critical temperature $J/\psi$, 
${\psi}'$, and $\chi_c$ become a mass triplet, and $D$ and $D^*$ ($D_s$ and 
$D^*_s$) become degenerate in mass. These are also true when the central
spin-independent potential and the smeared spin-spin interaction are used
in the Schr\"odinger equation to obtain temperature-dependent masses and
quark-antiquark relative-motion wave functions. The particular 
temperature dependence of the $J/\psi$ mass and radius causes the 
peak cross sections of the endothermic $\pi J/\psi$ and $\rho J/\psi$ reactions
to increase rapidly when temperature approaches the critical temperature. 
Even though ${\psi}'$ and $\chi_c$ are degenerate in mass, the cross 
sections of ${\psi}'$ and $\chi_c$ in collisions with a light meson are 
different, which is owed to the node of the  ${\psi}'$ wave function. The 
temperature dependence of the potential, the quark-antiquark relative-motion 
wave functions, and the meson masses leads to the temperature dependence of all
dissociation cross sections by the three factors, the difference $m_{q \bar c} 
+ m_{c \bar q} - m_{q \bar q} - m_{c \bar c}$, $\frac{1}{s}\frac{|\vec P'|}
{|\vec P|}$, and the transition amplitude. Some reactions are purely 
endothermic while the others become exothermic below certain temperatures. 
Parametrizations of unpolarized cross sections are given for 
$\pi J/\psi \to \bar{D}^{*} D$ or $\bar D D^*$, 
$\pi J/\psi \to\bar{D}^{*} D^{*}$, $\pi {\psi}' \to \bar{D}^* D$ or 
$\bar D D^*$, $\pi {\psi}' \to \bar{D}^* D^*$, $\pi \chi_{c} \to \bar{D}^* D$ 
or $\bar D D^*$, $\pi \chi_{c} \to \bar{D}^* D^*$, $\rho J/\psi \to \bar{D} D$,
$\rho J/\psi \to \bar{D}^* D$ or $\bar D D^*$, $\rho J/\psi \to \bar{D}^* D^*$,
$\rho {\psi}' \to \bar{D} D$, $\rho {\psi}' \to \bar{D}^* D$ or 
$\bar D D^*$, $\rho {\psi}' \to \bar{D}^* D^*$, $\rho \chi_{c} \to \bar{D} D$, 
$\rho \chi_{c} \to \bar{D}^* D$ or $\bar D D^*$, and 
$\rho \chi_{c} \to \bar{D}^* D^*$. The parametrizations at the five 
temperatures, $T/T_{\rm c}=$ 0.65, 0.75, 0.85, 0.9, 0.95, can be used in the 
procedure to yield cross sections at any temperature between 0.65$T_{\rm c}$ 
and $T_{\rm c}$.

\vspace{0.5cm}
\leftline{\bf Acknowledgements}
\vspace{0.5cm}
This work was supported by the
National Natural Science Foundation of China under Grant No. 11175111.

\newpage

\begin{figure}[htbp]
\centering
\includegraphics[scale=0.8]{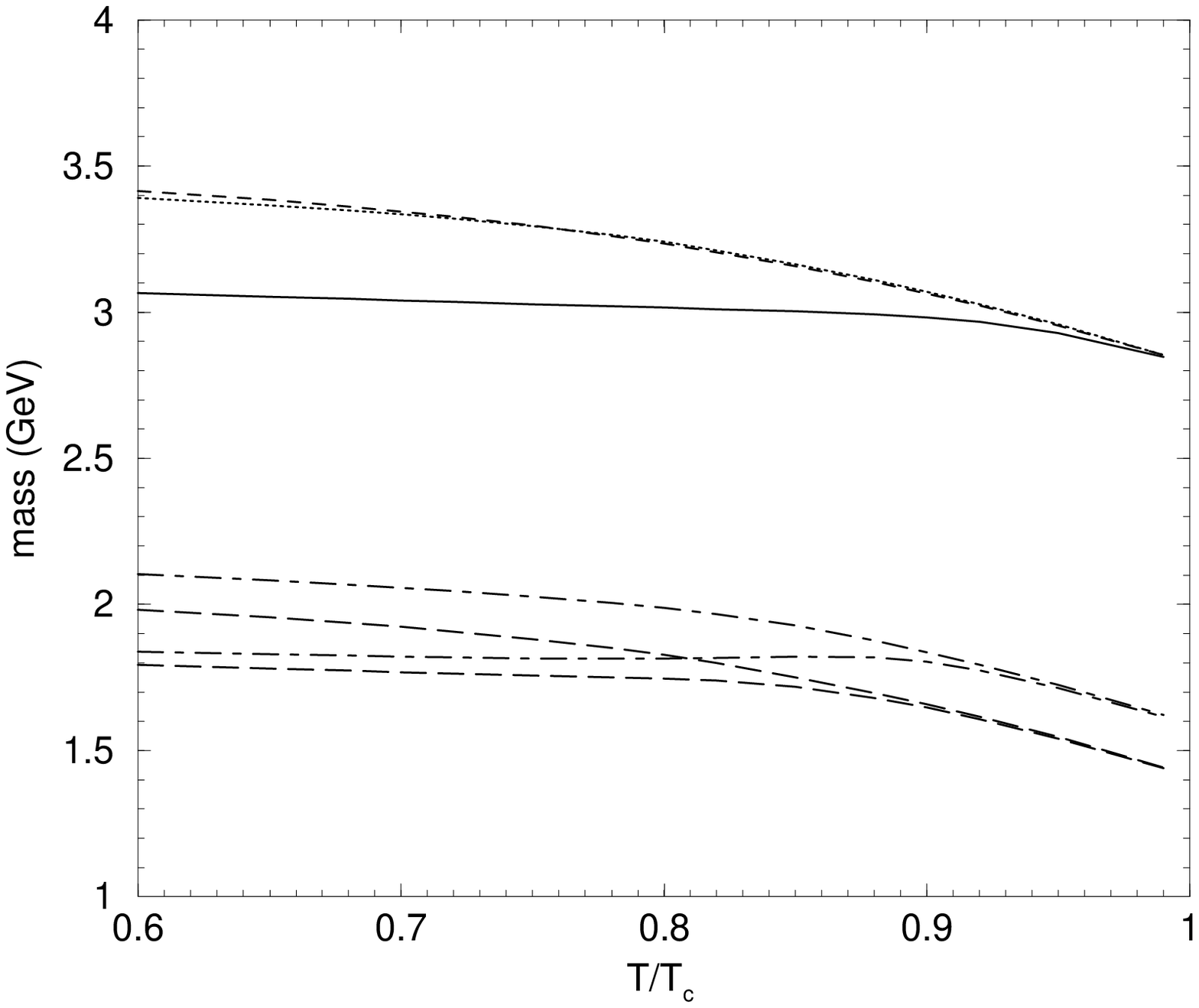}%
\caption{Meson masses as functions of $T/T_{\rm c}$. The masses of $J/\psi$, 
${\psi}'$, $\chi_c$, $D$, $D^*$, $D_s$ and $D^*_s$ are shown by the solid, 
dashed, dotted, lower long dashed, upper long dashed, lower dot-dashed and 
upper dot-dashed curves, respectively.}
\label{fig1}
\end{figure}

\newpage

\begin{figure}[htbp]
\centering
\includegraphics[scale=0.8]{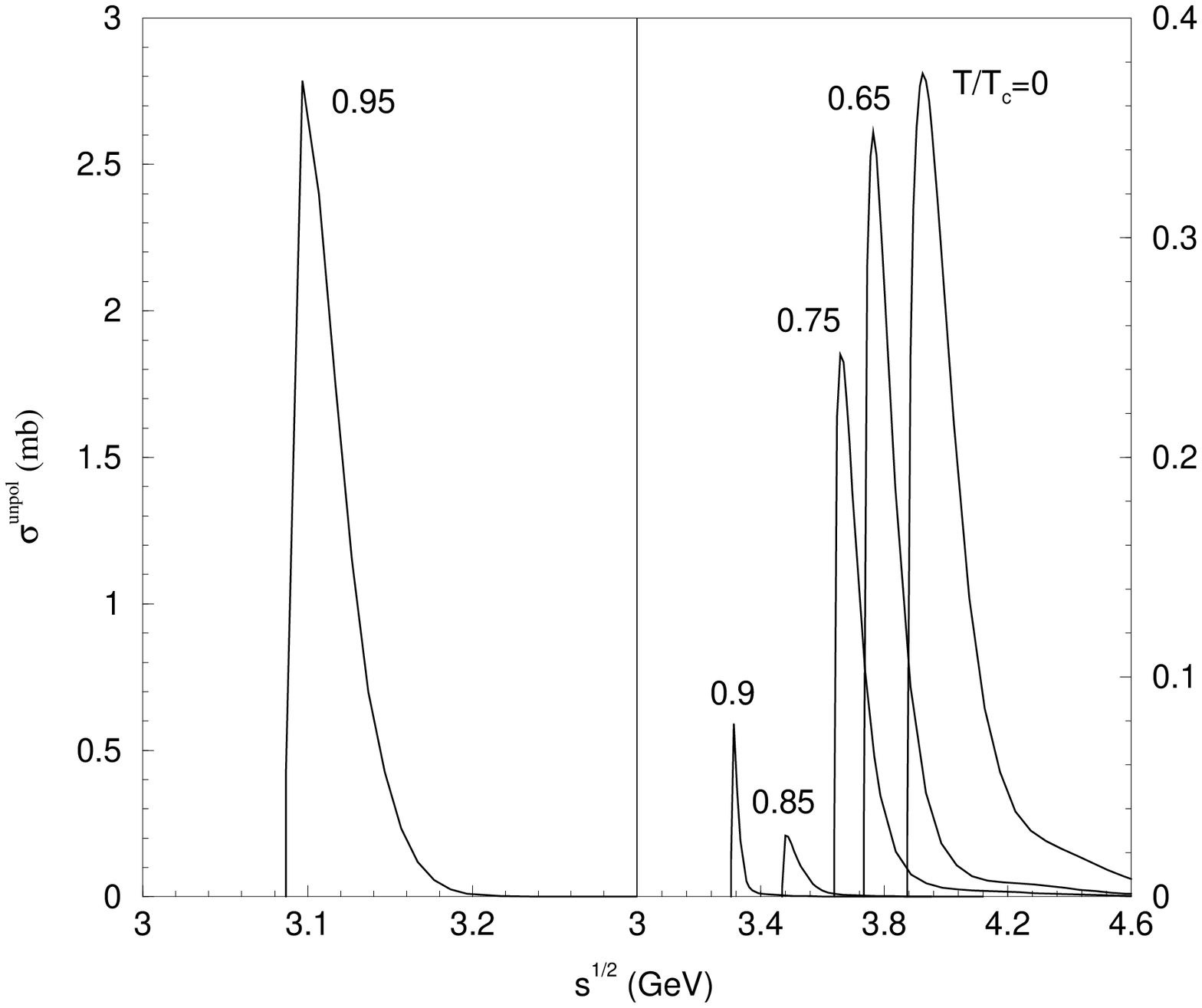}%
\caption{Cross sections for $\pi J/\psi \to \bar{D}^* D$ or $\bar D D^*$ at 
various temperatures.}
\label{fig2}
\end{figure}

\newpage

\begin{figure}[htbp]
\centering
\includegraphics[scale=0.8]{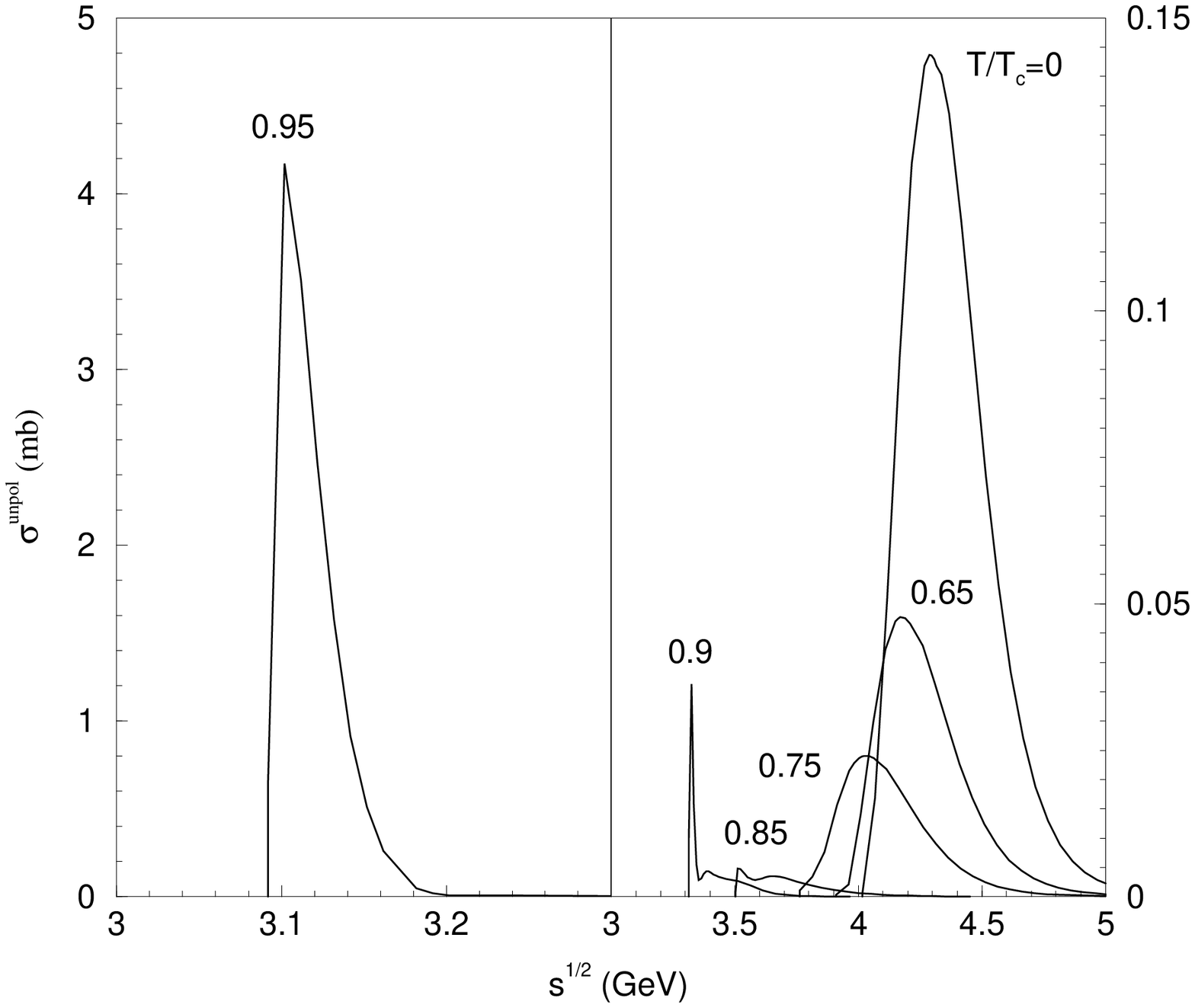}%
\caption{Cross sections for $\pi J/\psi \to \bar{D}^* D^*$ at various 
temperatures.}
\label{fig3}
\end{figure}

\newpage

\begin{figure}[htbp]
\centering
\includegraphics[scale=0.8]{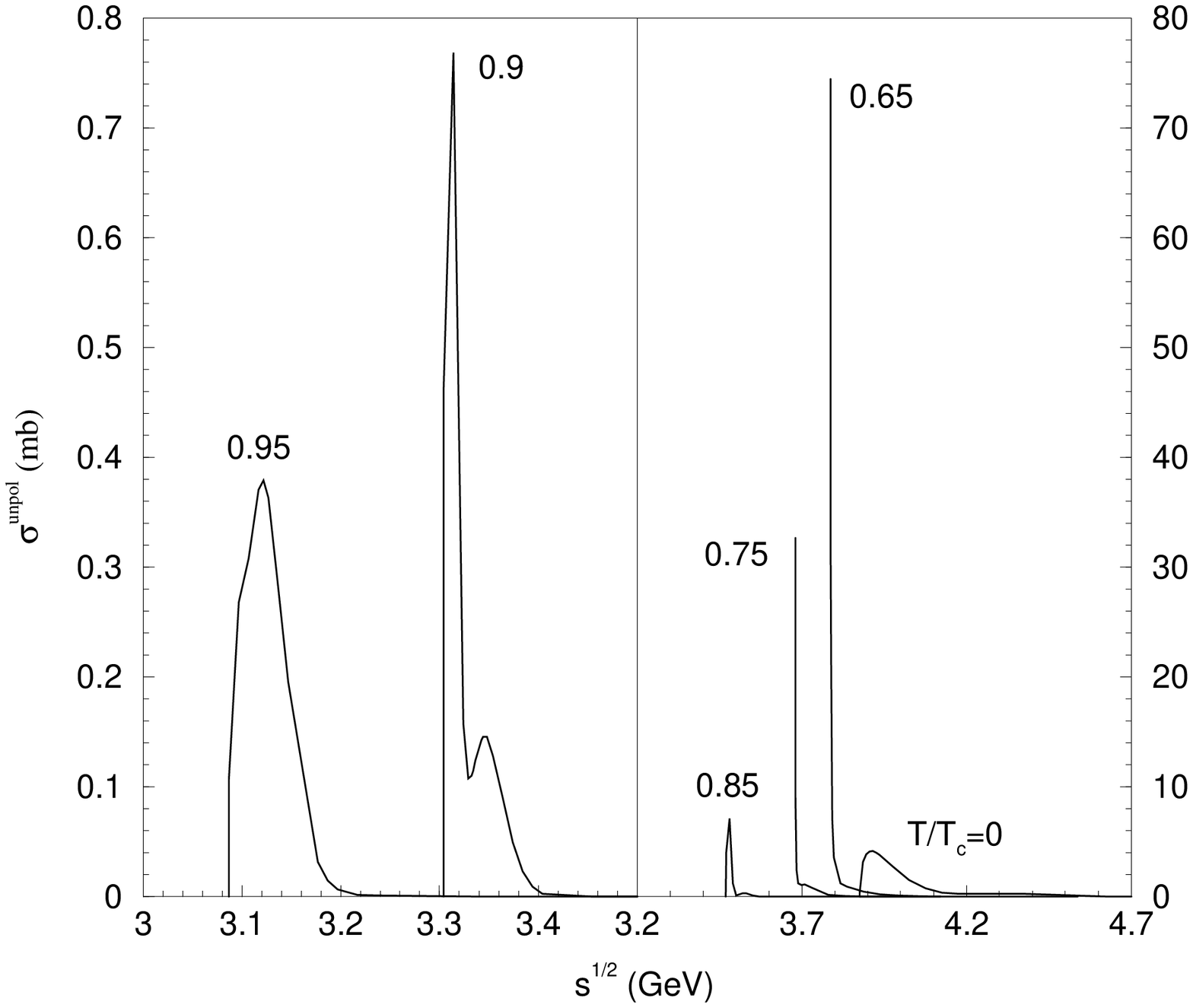}%
\caption{Cross sections for $\pi {\psi}' \to \bar{D}^* D$ or $\bar D D^*$ at 
various temperatures.}
\label{fig4}
\end{figure}

\newpage

\begin{figure}[htbp]
\centering
\includegraphics[scale=0.8]{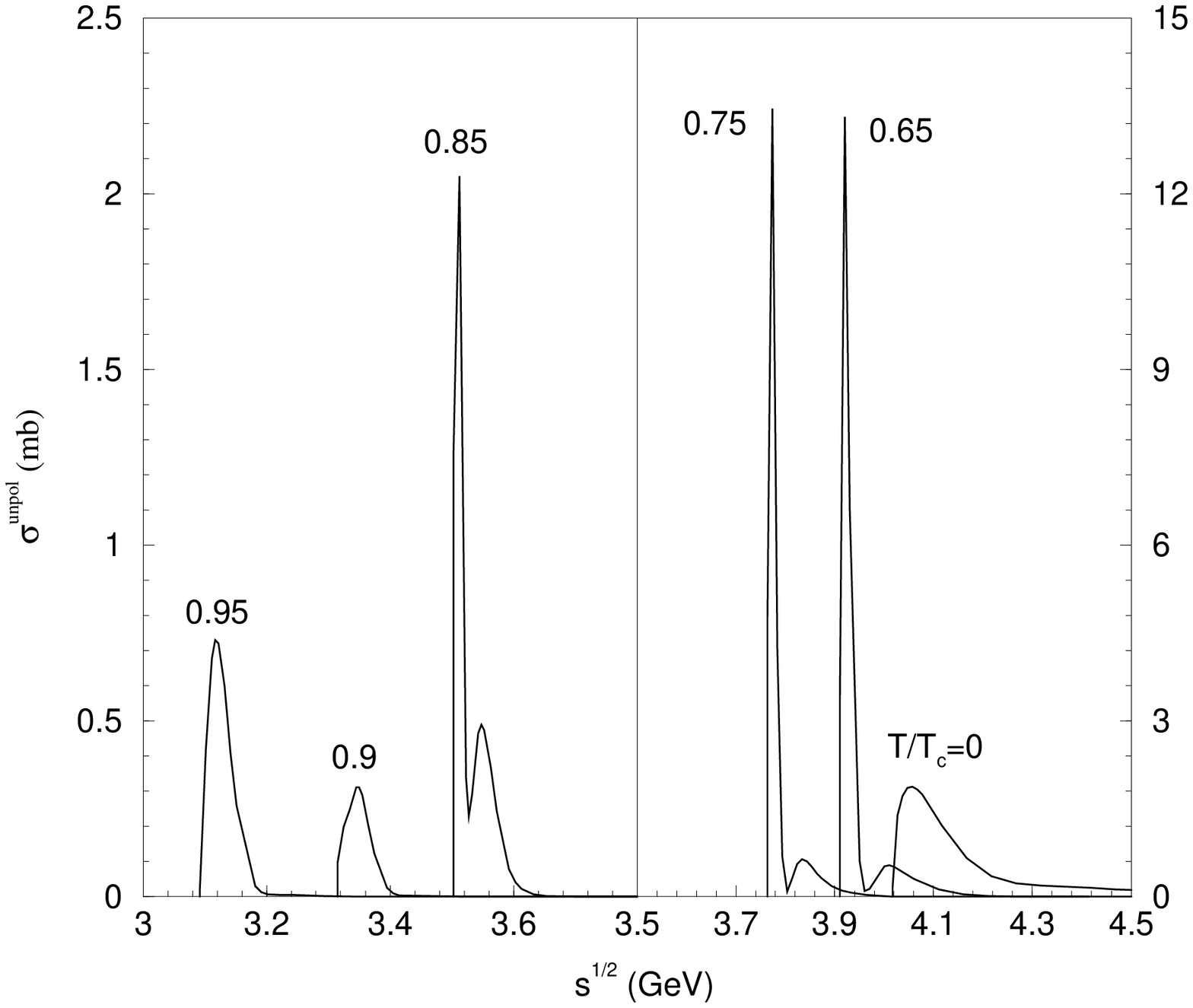}%
\caption{Cross sections for $\pi {\psi}' \to \bar{D}^* D^*$ at various 
temperatures.}
\label{fig5}
\end{figure}

\newpage

\begin{figure}[htbp]
\centering
\includegraphics[scale=0.8]{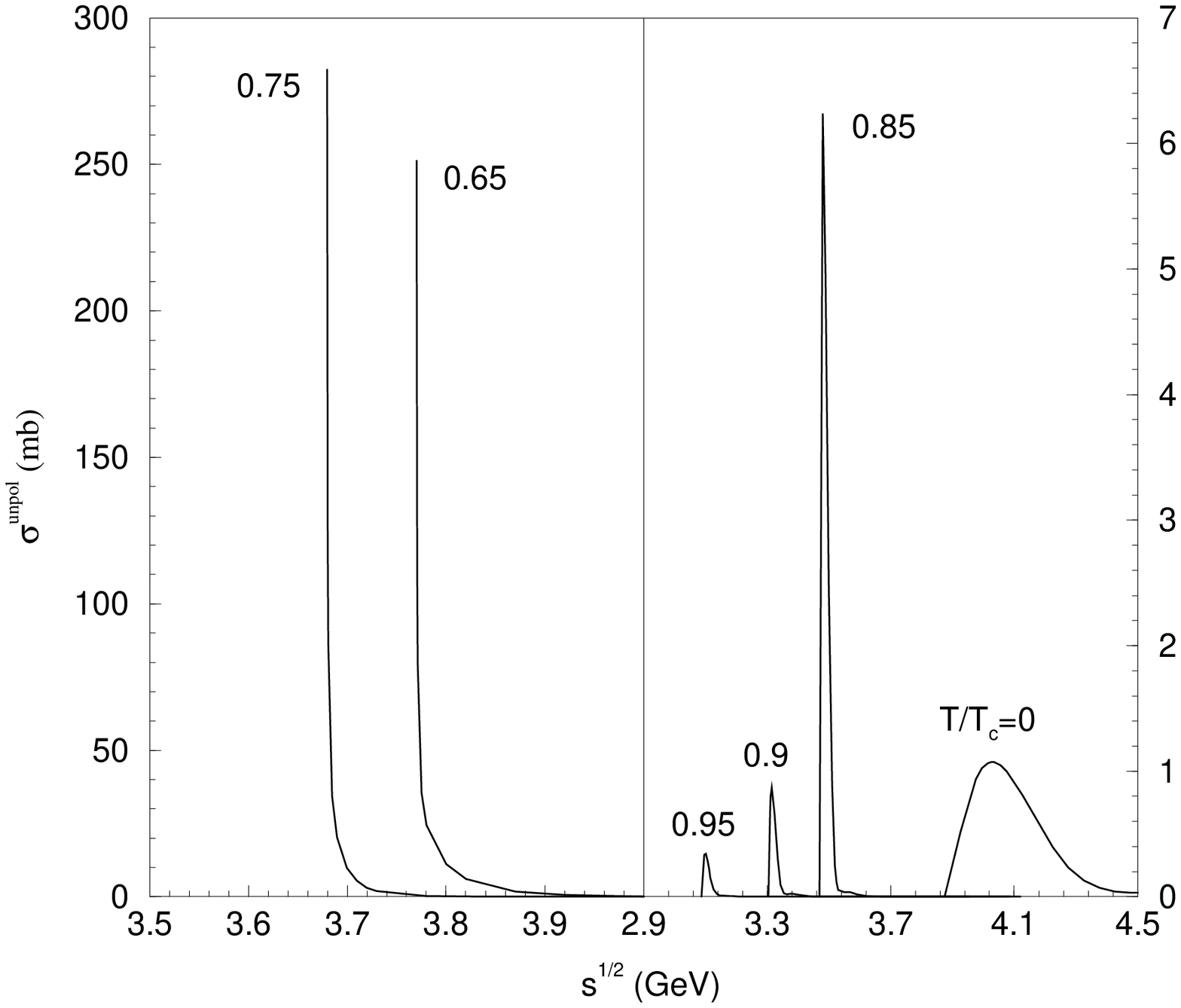}%
\caption{Cross sections for $\pi \chi_c \to \bar{D}^* D$ or $\bar D D^*$ at 
various temperatures.}
\label{fig6}
\end{figure}

\newpage

\begin{figure}[htbp]
\centering
\includegraphics[scale=0.8]{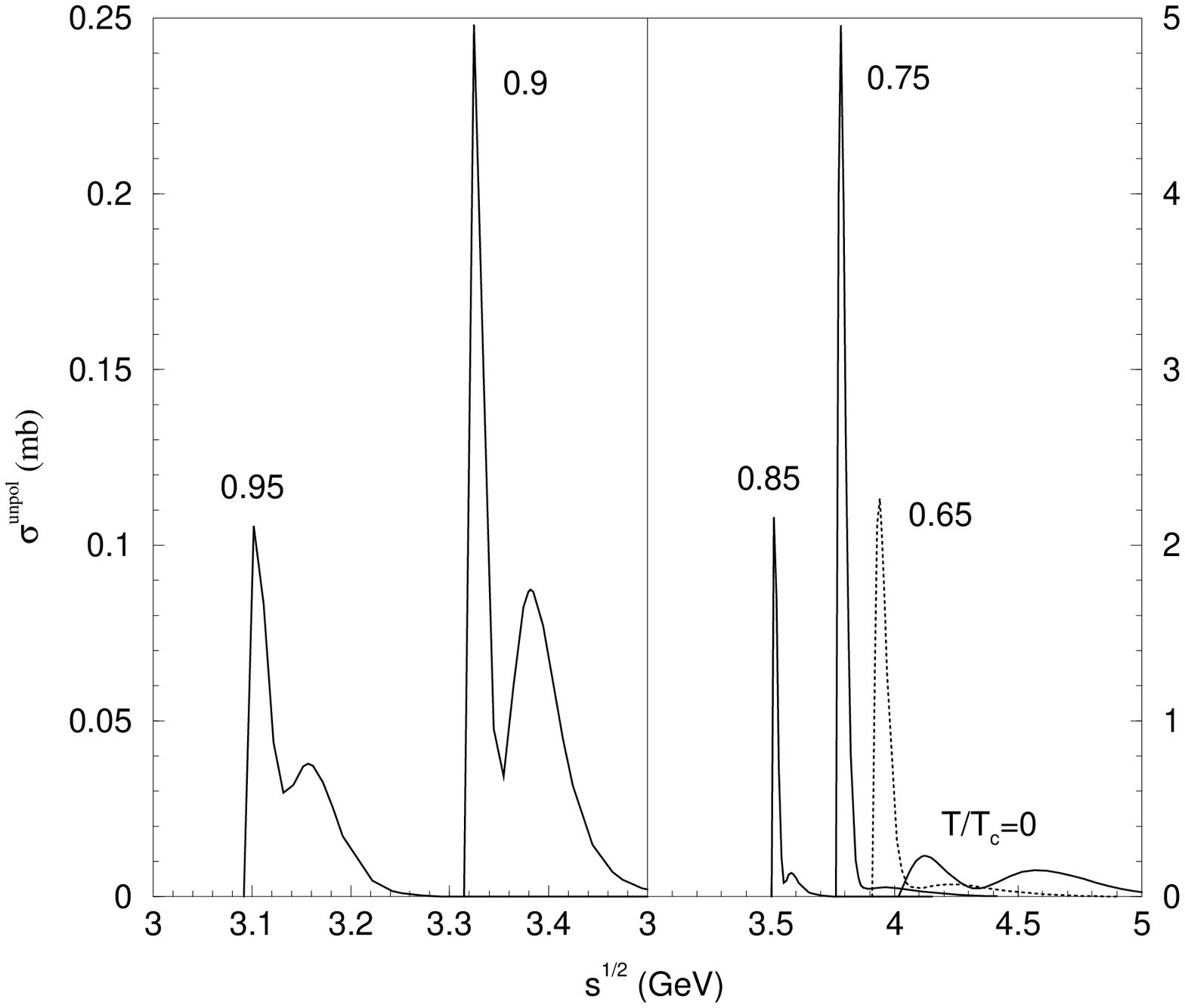}%
\caption{Cross sections for $\pi \chi_c \to \bar{D}^* D^*$ at various 
temperatures.}
\label{fig7}
\end{figure}

\newpage

\begin{figure}[htbp]
\centering
\includegraphics[scale=0.8]{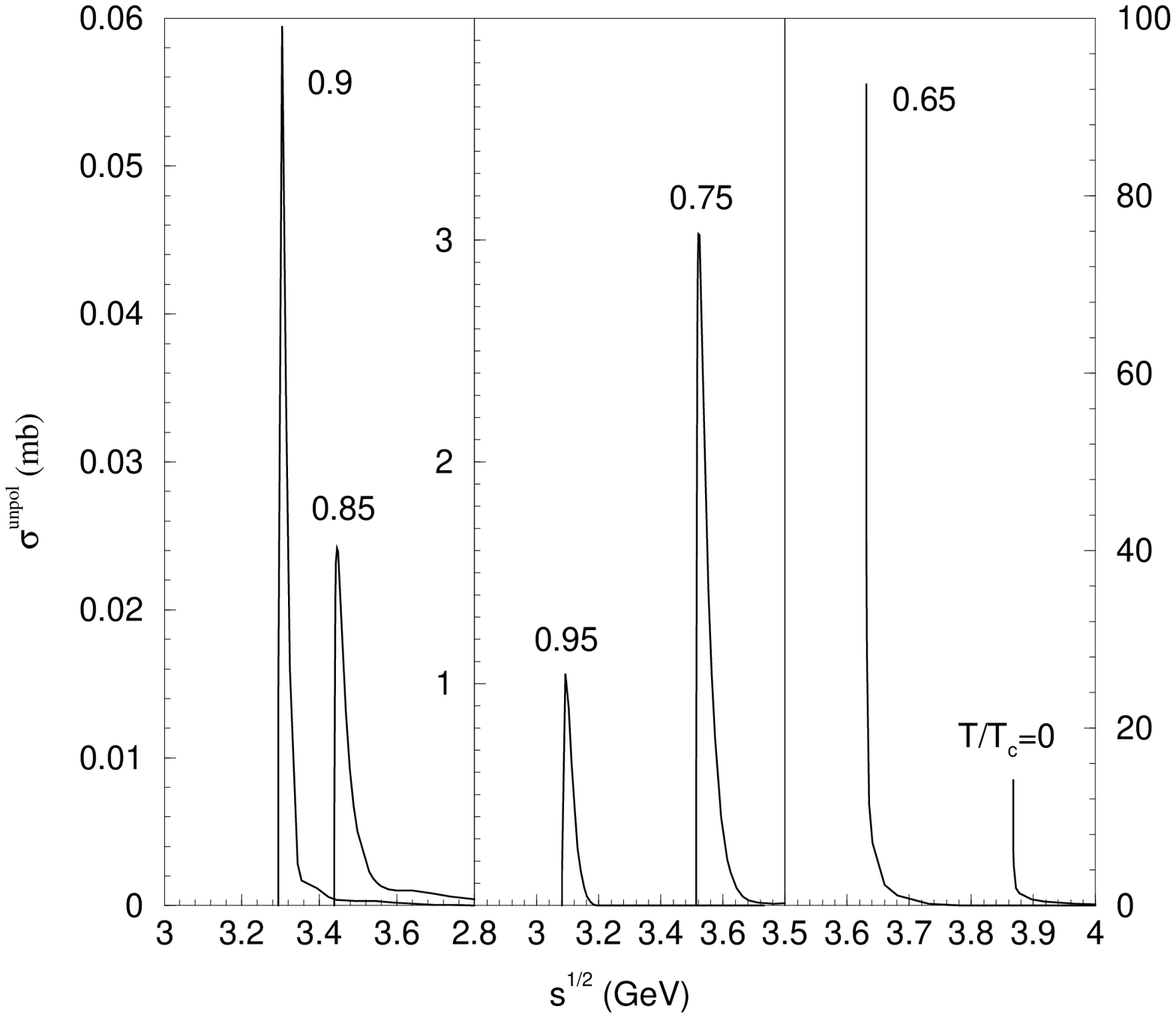}%
\caption{Cross sections for $\rho J/\psi \to \bar{D} D$ at various 
temperatures.}
\label{fig8}
\end{figure}

\newpage

\begin{figure}[htbp]
\centering
\includegraphics[scale=0.8]{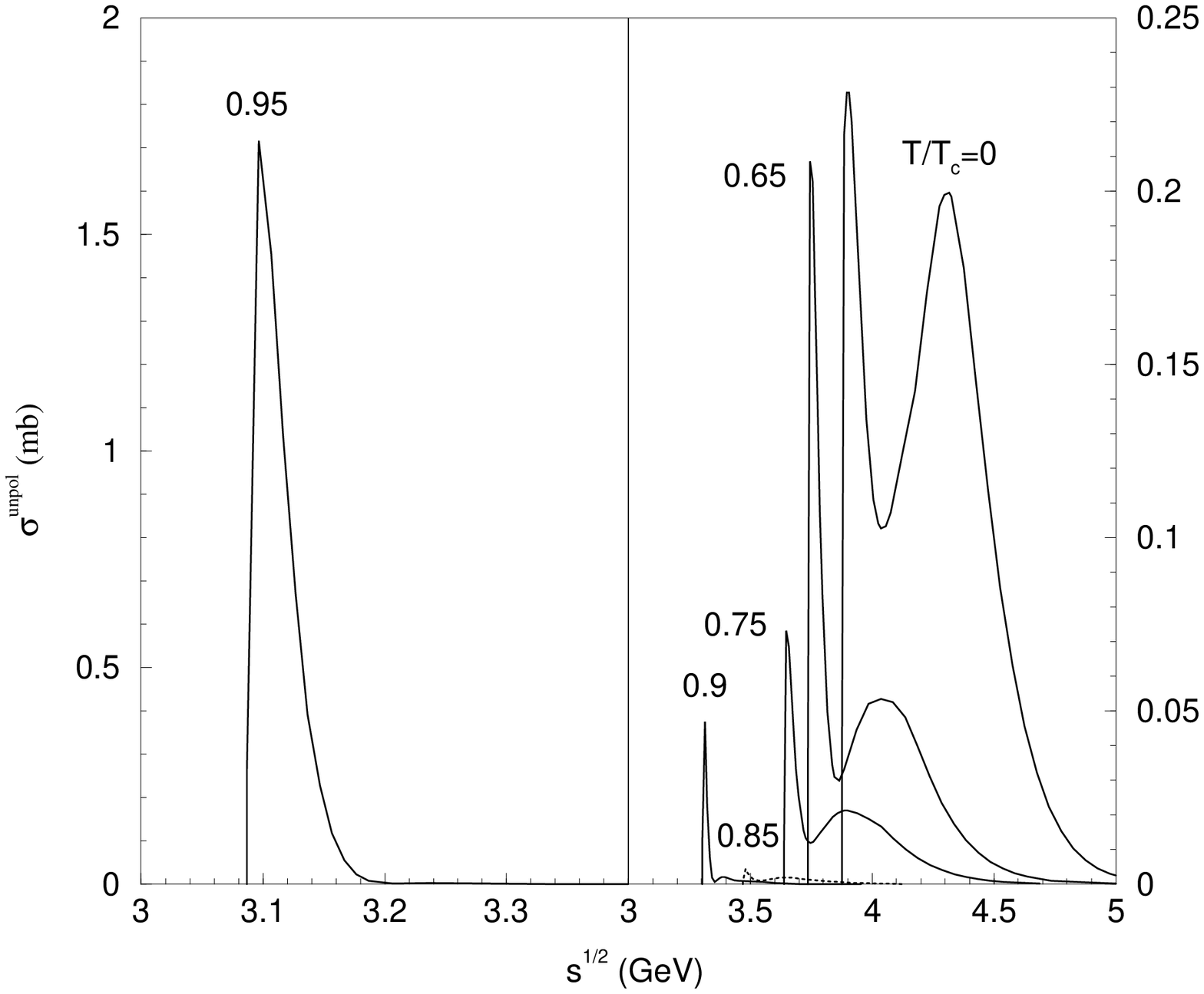}%
\caption{Cross sections for $\rho J/\psi \to \bar{D}^* D$ or $\bar D D^*$ at 
various temperatures.}
\label{fig9}
\end{figure}

\newpage

\begin{figure}[htbp]
\centering
\includegraphics[scale=0.8]{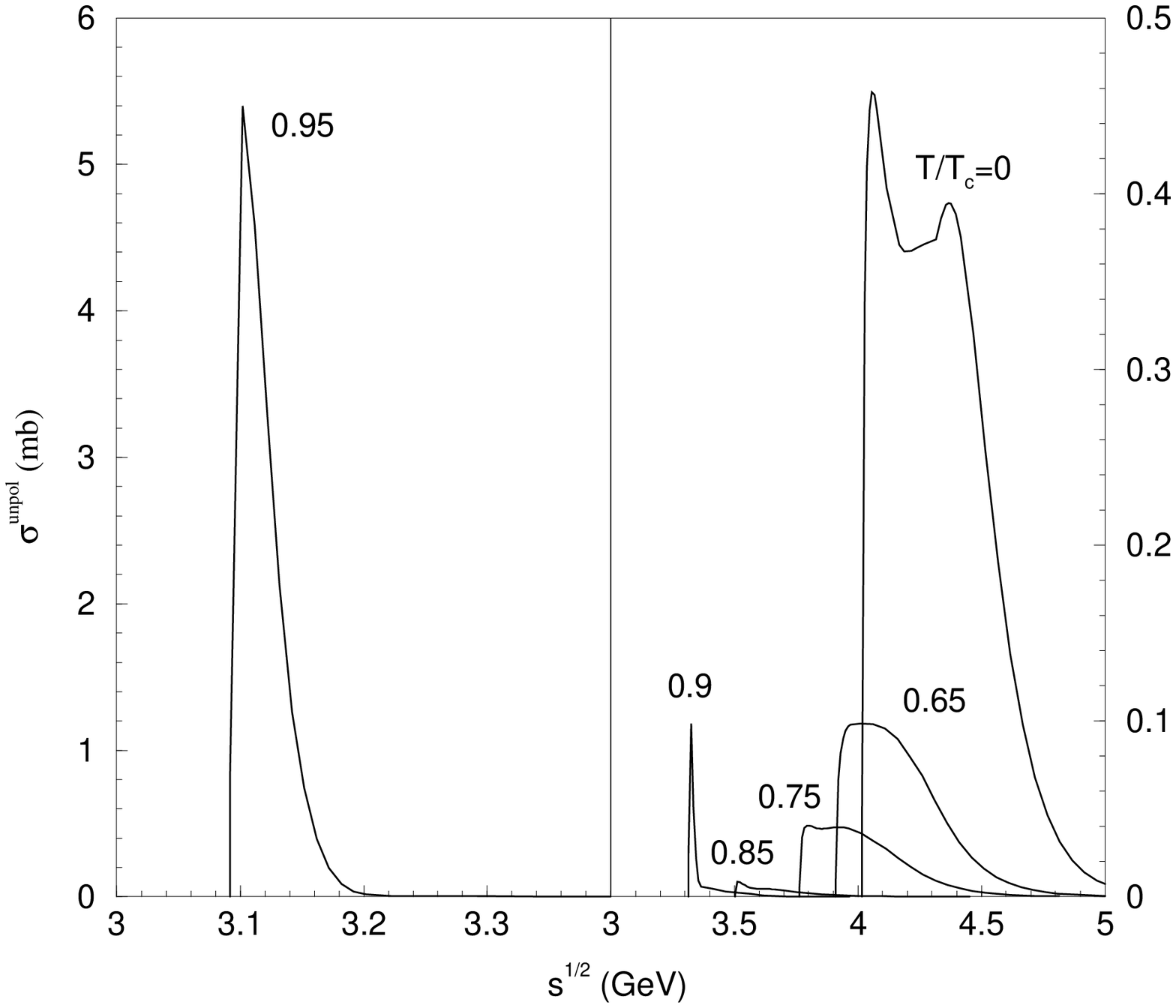}%
\caption{Cross sections for $\rho J/\psi \to \bar{D}^* D^*$ at various 
temperatures.}
\label{fig10}
\end{figure}

\newpage

\begin{figure}[htbp]
\centering
\includegraphics[scale=0.8]{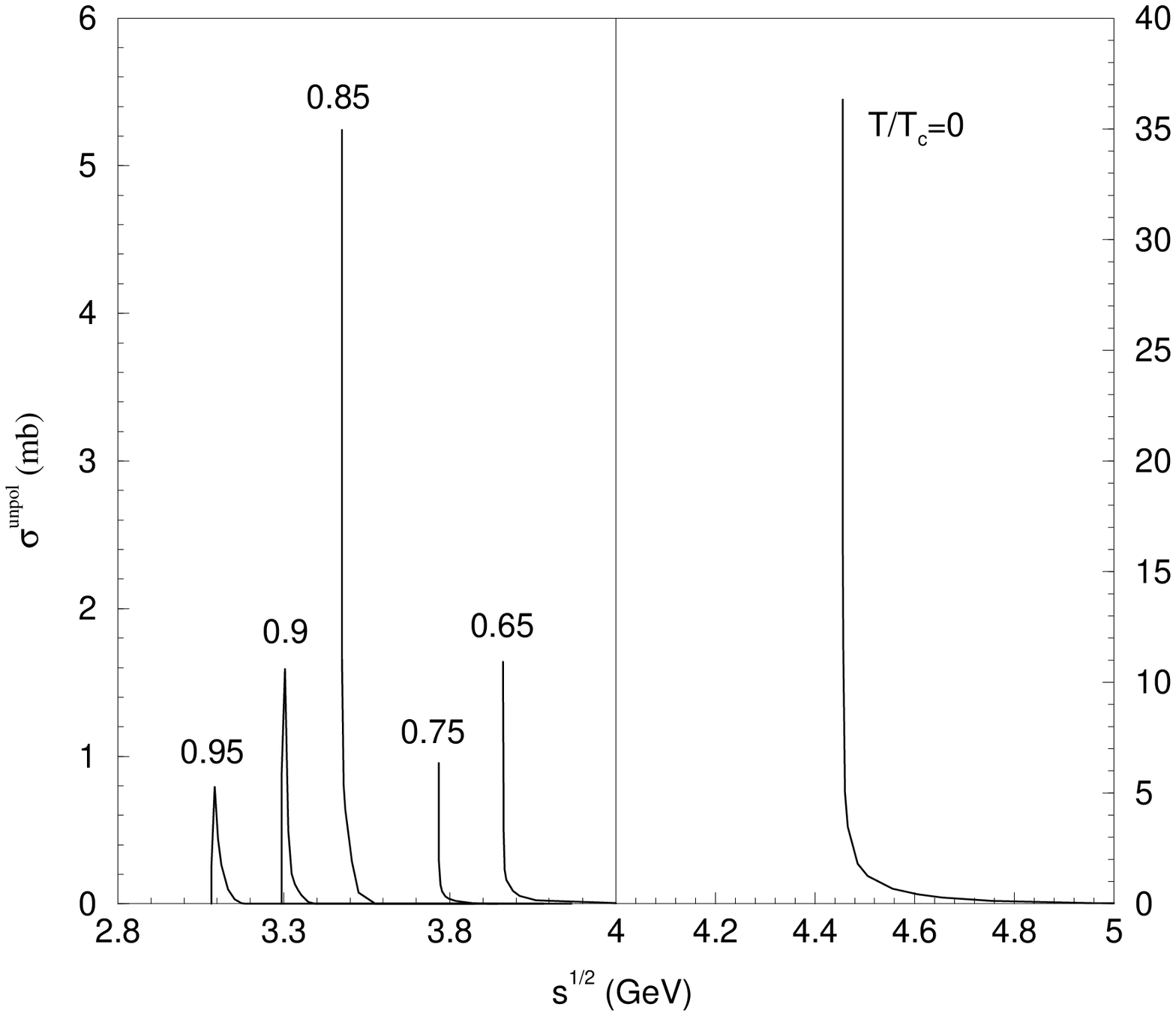}%
\caption{Cross sections for $\rho {\psi}' \to \bar{D} D$ at various 
temperatures.}
\label{fig11}
\end{figure}

\newpage

\begin{figure}[htbp]
\centering
\includegraphics[scale=0.8]{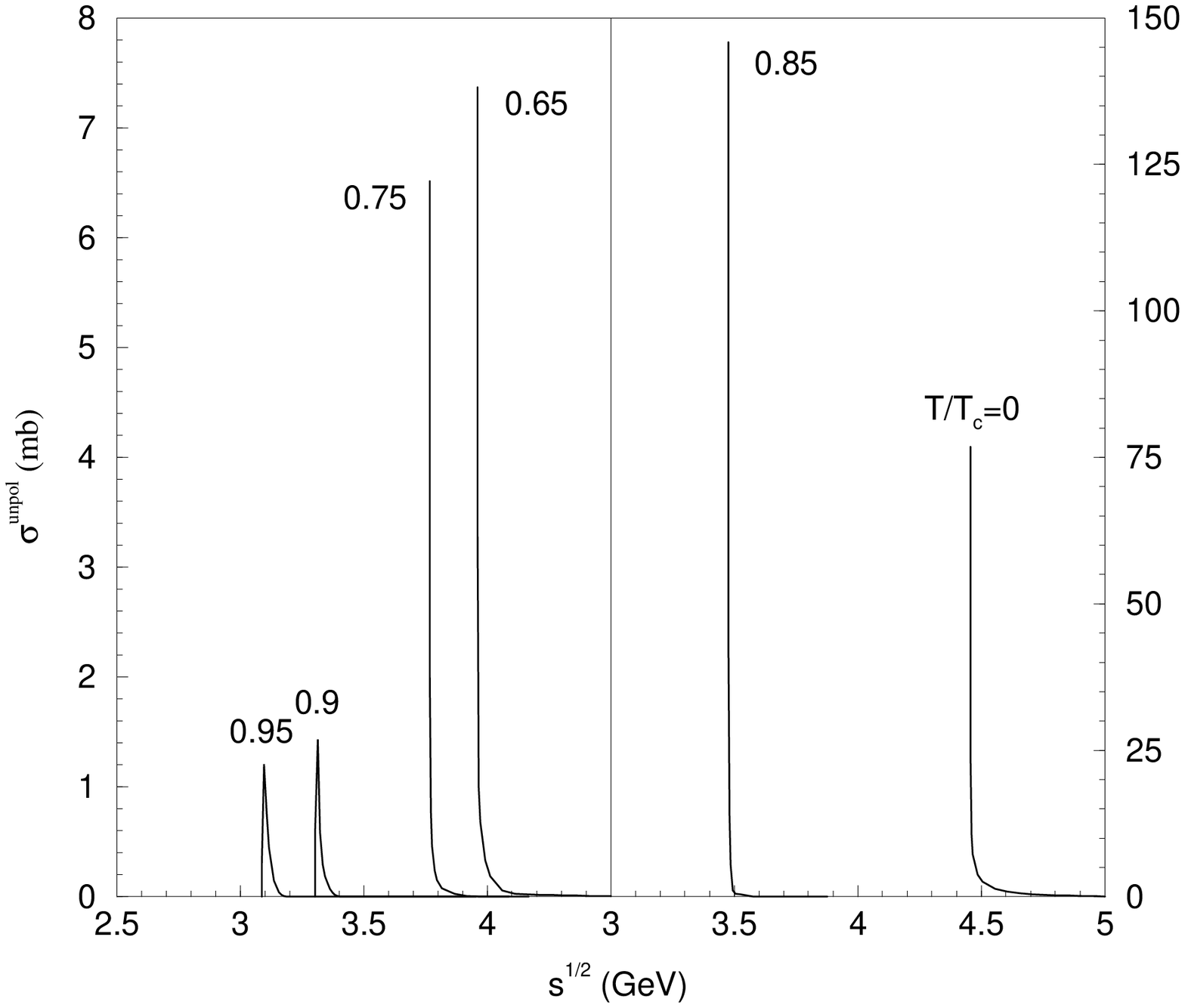}%
\caption{Cross sections for $\rho {\psi}' \to \bar{D}^* D$ or $\bar D D^*$ at 
various temperatures.}
\label{fig12}
\end{figure}

\newpage

\begin{figure}[htbp]
\centering
\includegraphics[scale=0.8]{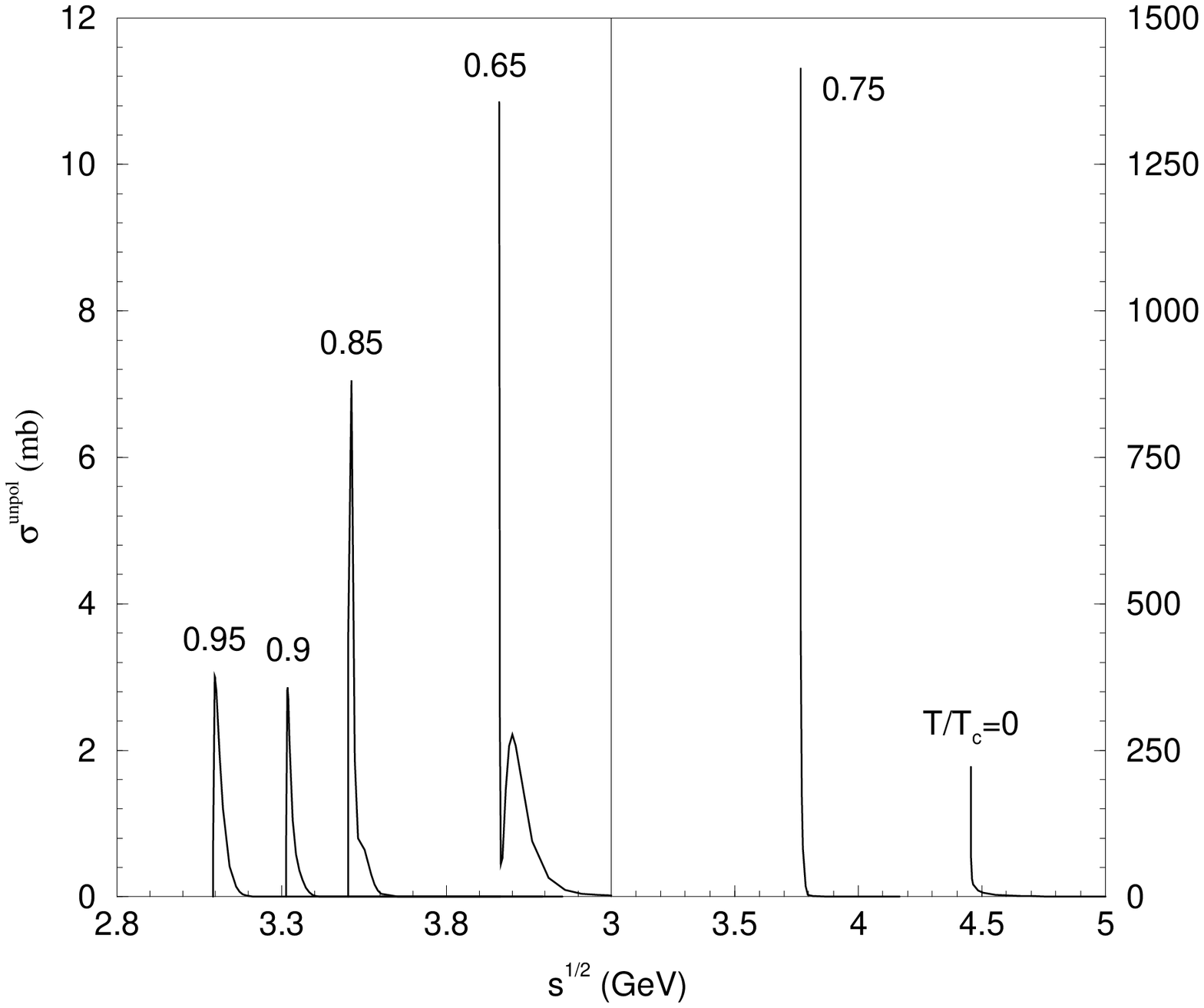}%
\caption{Cross sections for $\rho {\psi}' \to \bar{D}^* D^*$ at various 
temperatures.}
\label{fig13}
\end{figure}

\newpage

\begin{figure}[htbp]
\centering
\includegraphics[scale=0.8]{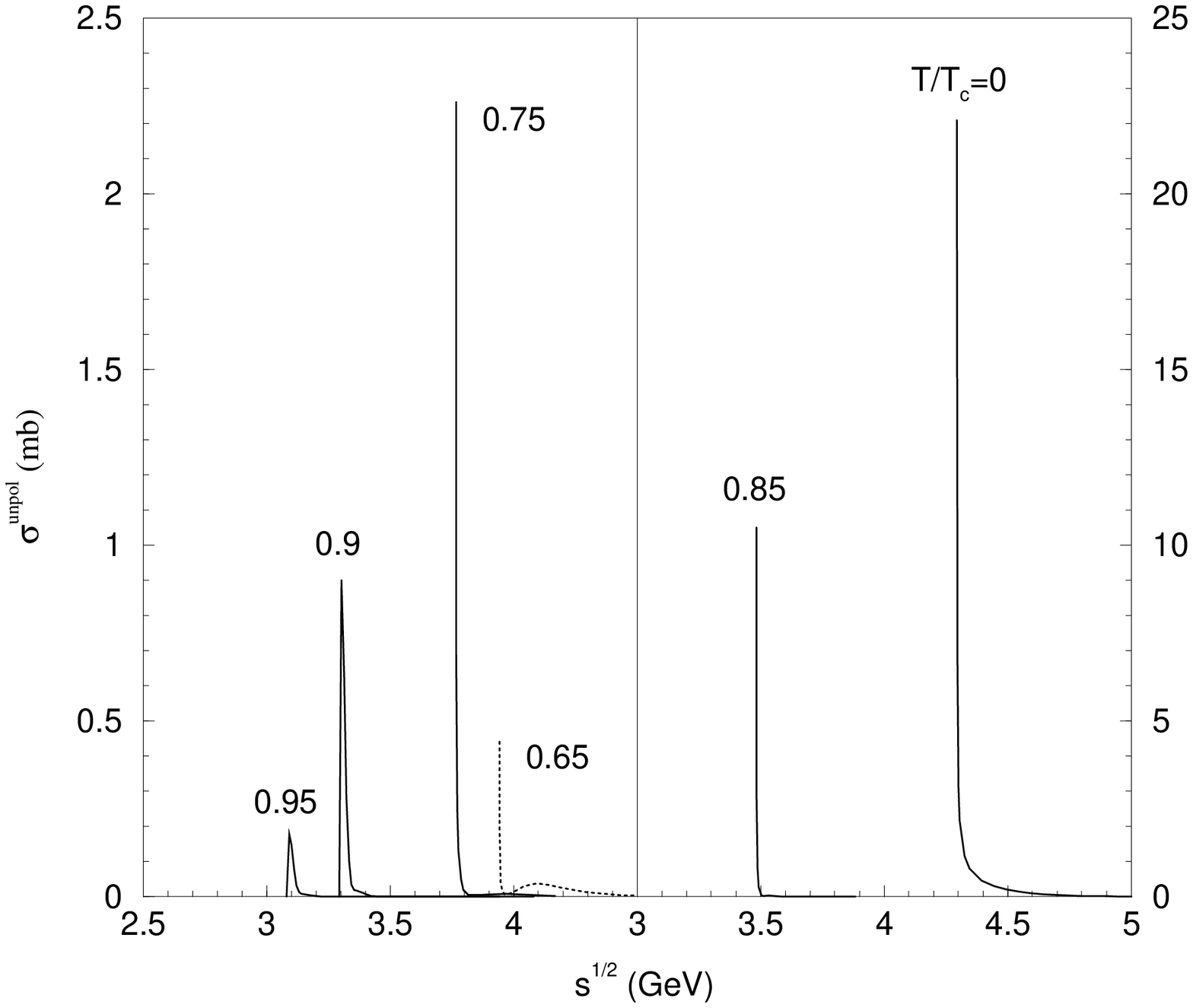}%
\caption{Cross sections for $\rho \chi_c \to \bar{D} D$ at various 
temperatures.}
\label{fig14}
\end{figure}

\newpage

\begin{figure}[htbp]
\centering
\includegraphics[scale=0.8]{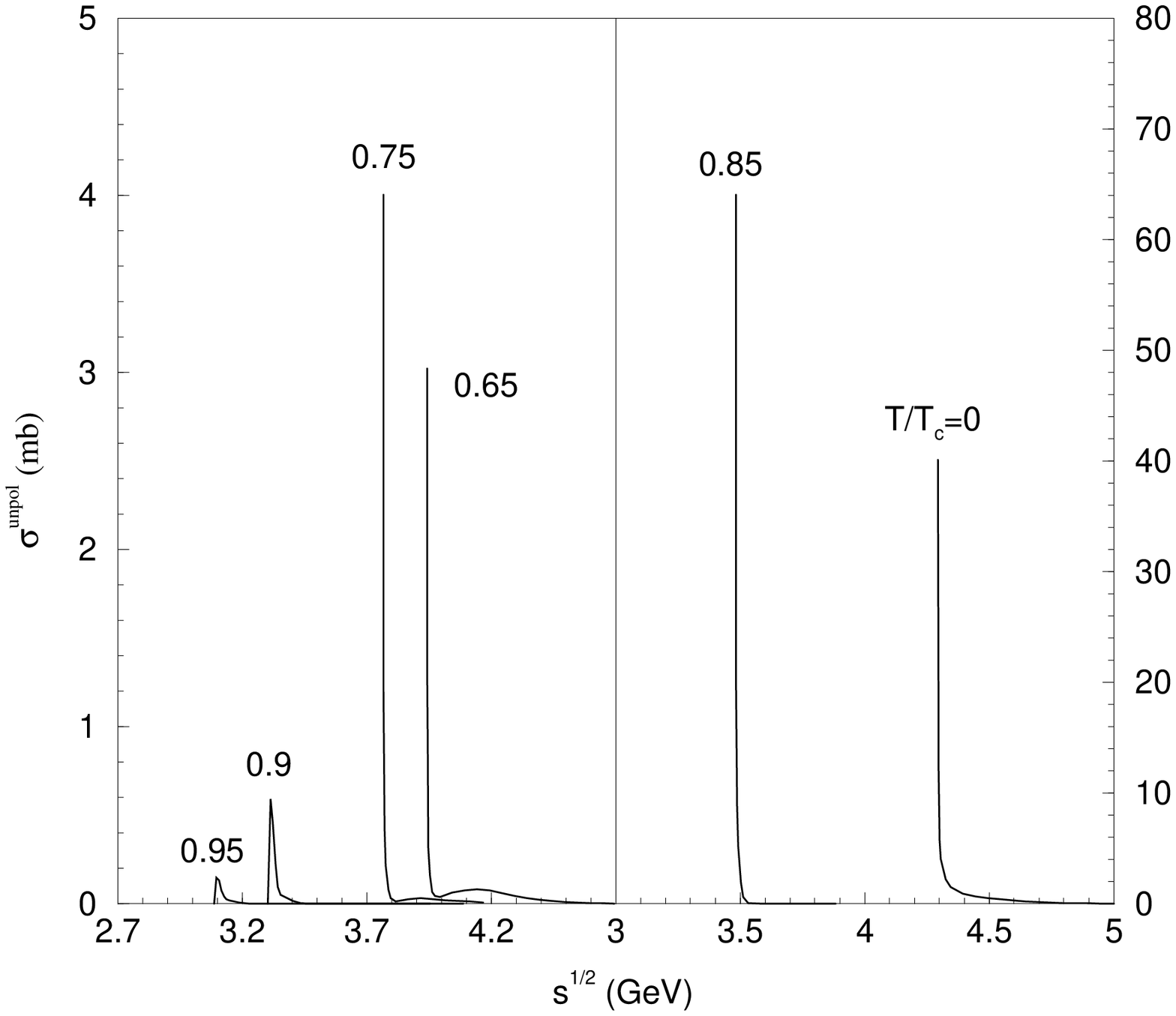}%
\caption{Cross sections for $\rho \chi_c \to \bar{D}^* D$ or $\bar D D^*$ at 
various temperatures.}
\label{fig15}
\end{figure}

\newpage

\begin{figure}[htbp]
\centering
\includegraphics[scale=0.8]{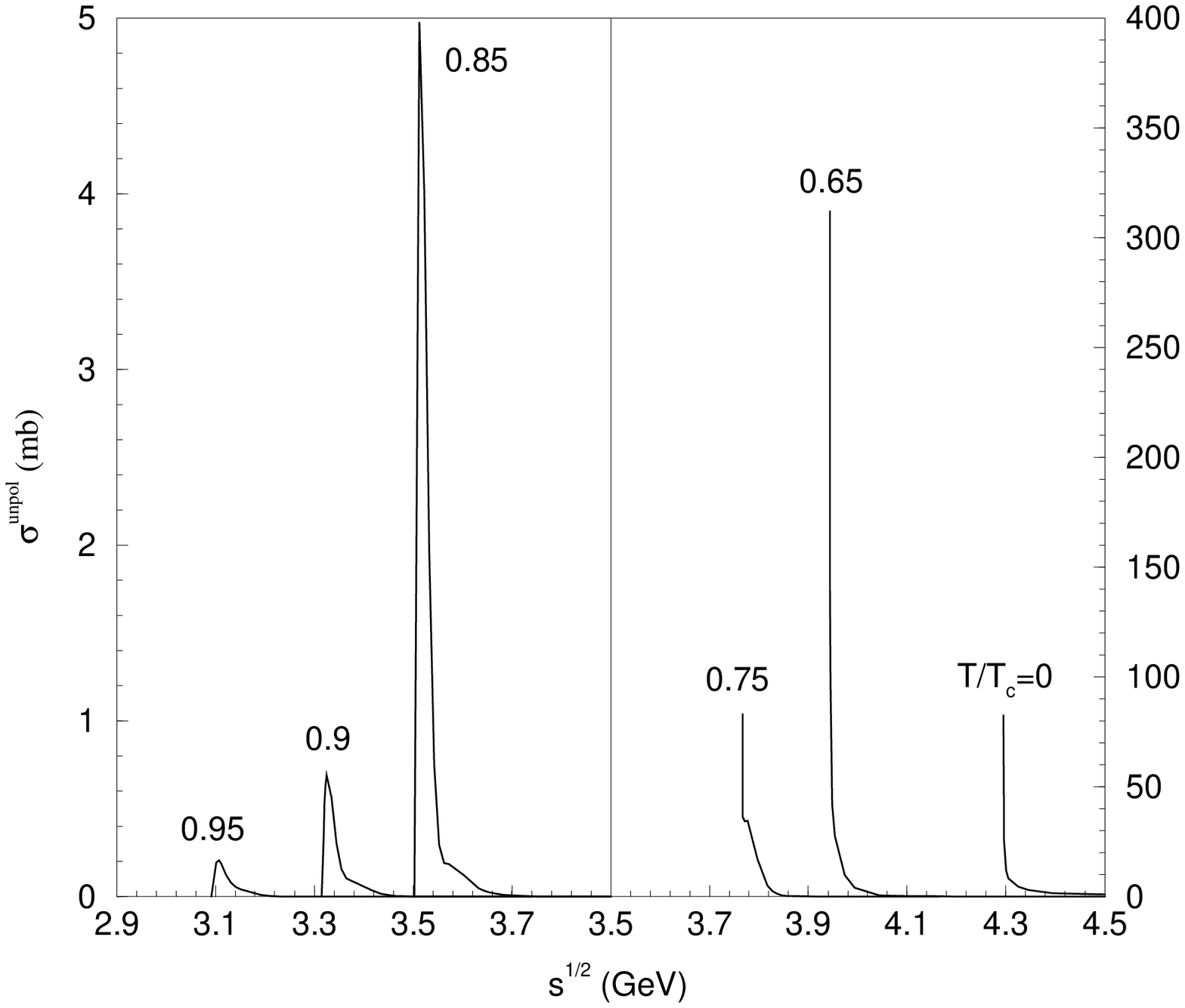}%
\caption{Cross sections for $\rho \chi_c \to \bar{D}^* D^*$ at various 
temperatures.}
\label{fig16}
\end{figure}

\newpage

\begin{figure}[htbp]
\centering
\includegraphics[scale=0.8]{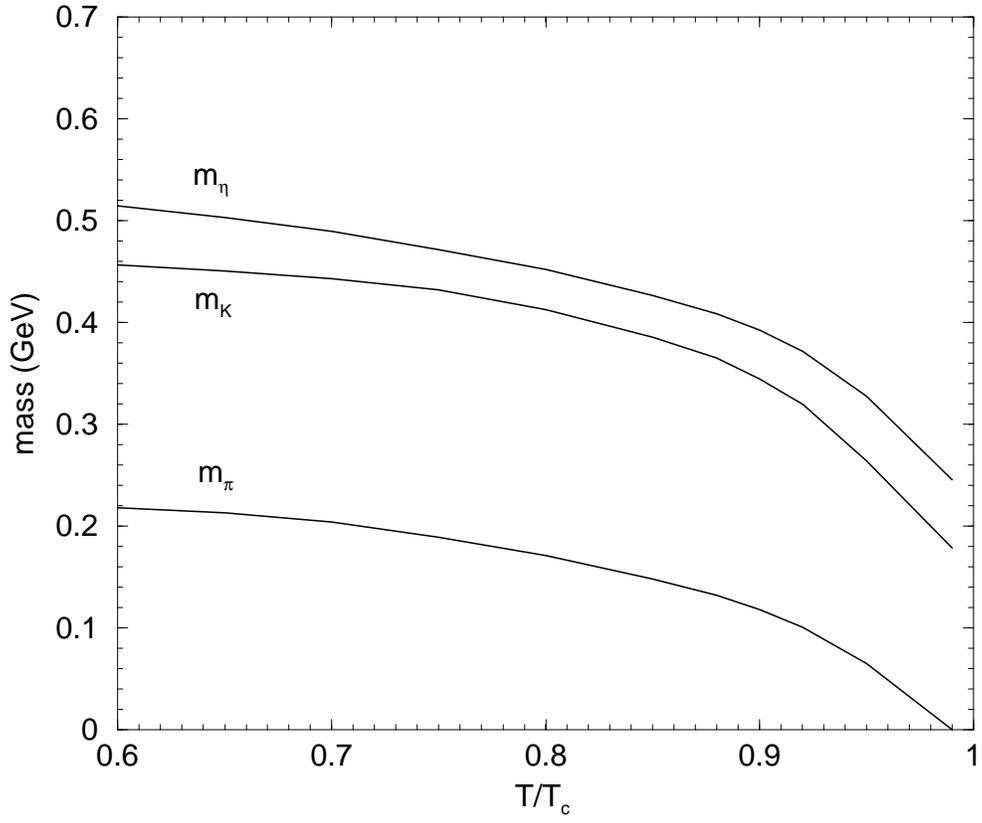}%
\caption{$\pi$, $K$ and $\eta$ masses as functions of $T/T_{\rm c}$.}
\label{fig17}
\end{figure}

\newpage

\begin{figure}[htbp]
\centering
\includegraphics[scale=0.8]{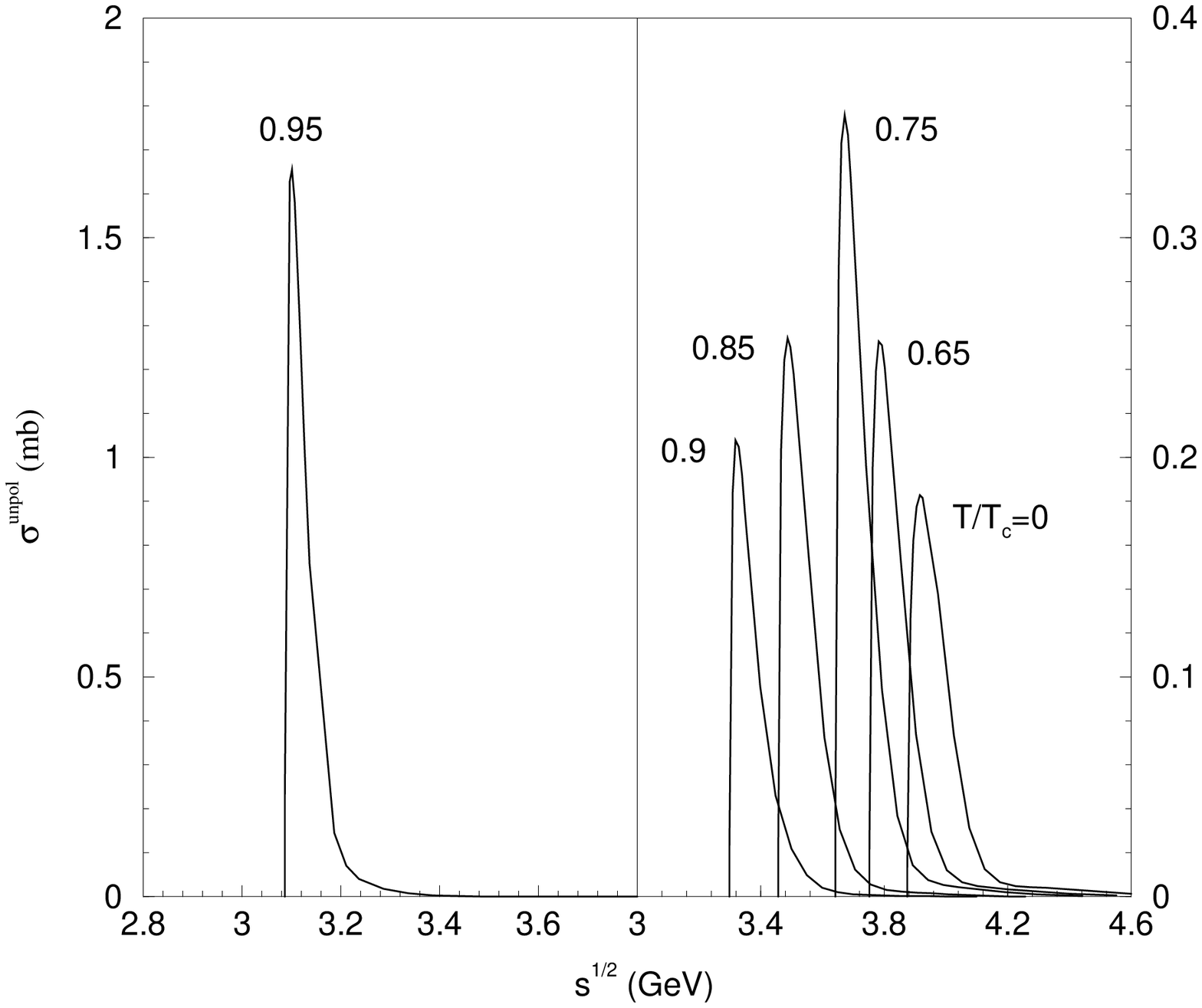}%
\caption{Cross sections for $\pi J/\psi \to \bar{D}^* D$ or $\bar D D^*$ at 
various temperatures.}
\label{fig18}
\end{figure}

\newpage

\begin{figure}[htbp]
\centering
\includegraphics[scale=0.8]{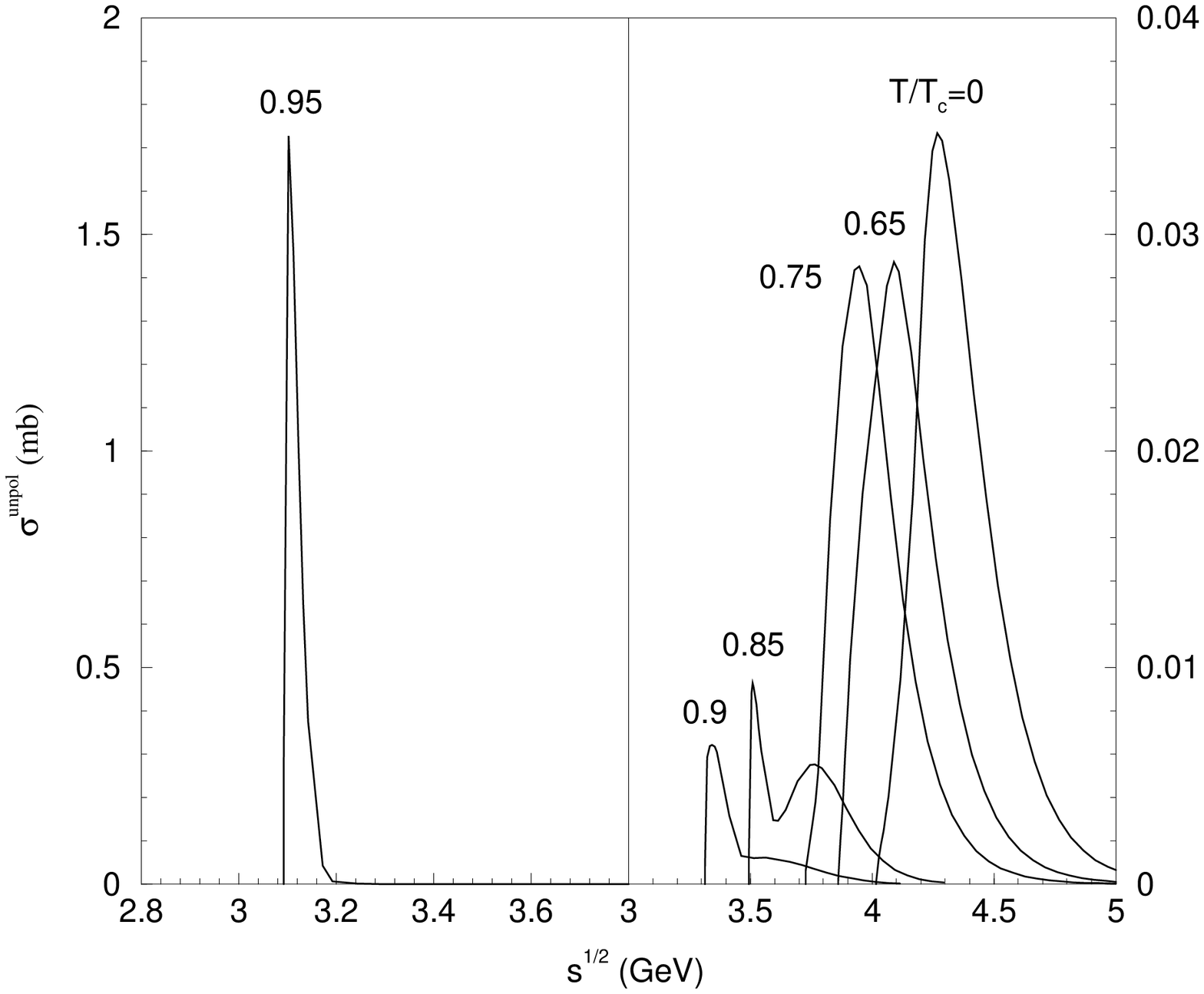}%
\caption{Cross sections for $\pi J/\psi \to \bar{D}^* D^*$ at various 
temperatures.}
\label{fig19}
\end{figure}

\newpage

\begin{figure}[htbp]
\centering
\includegraphics[scale=0.8]{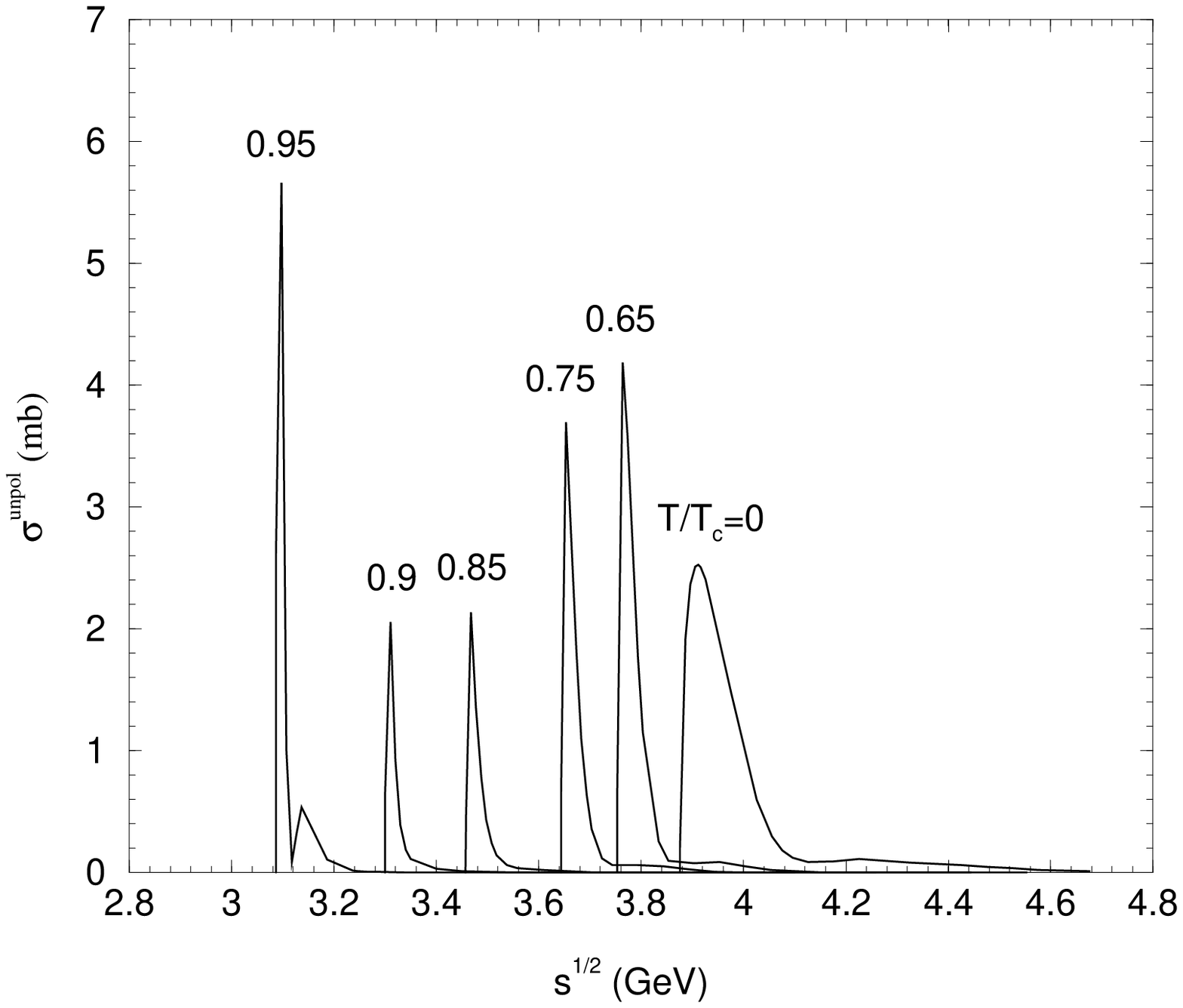}%
\caption{Cross sections for $\pi {\psi}' \to \bar{D}^* D$ or $\bar D D^*$ at 
various temperatures.}
\label{fig20}
\end{figure}

\newpage

\begin{figure}[htbp]
\centering
\includegraphics[scale=0.8]{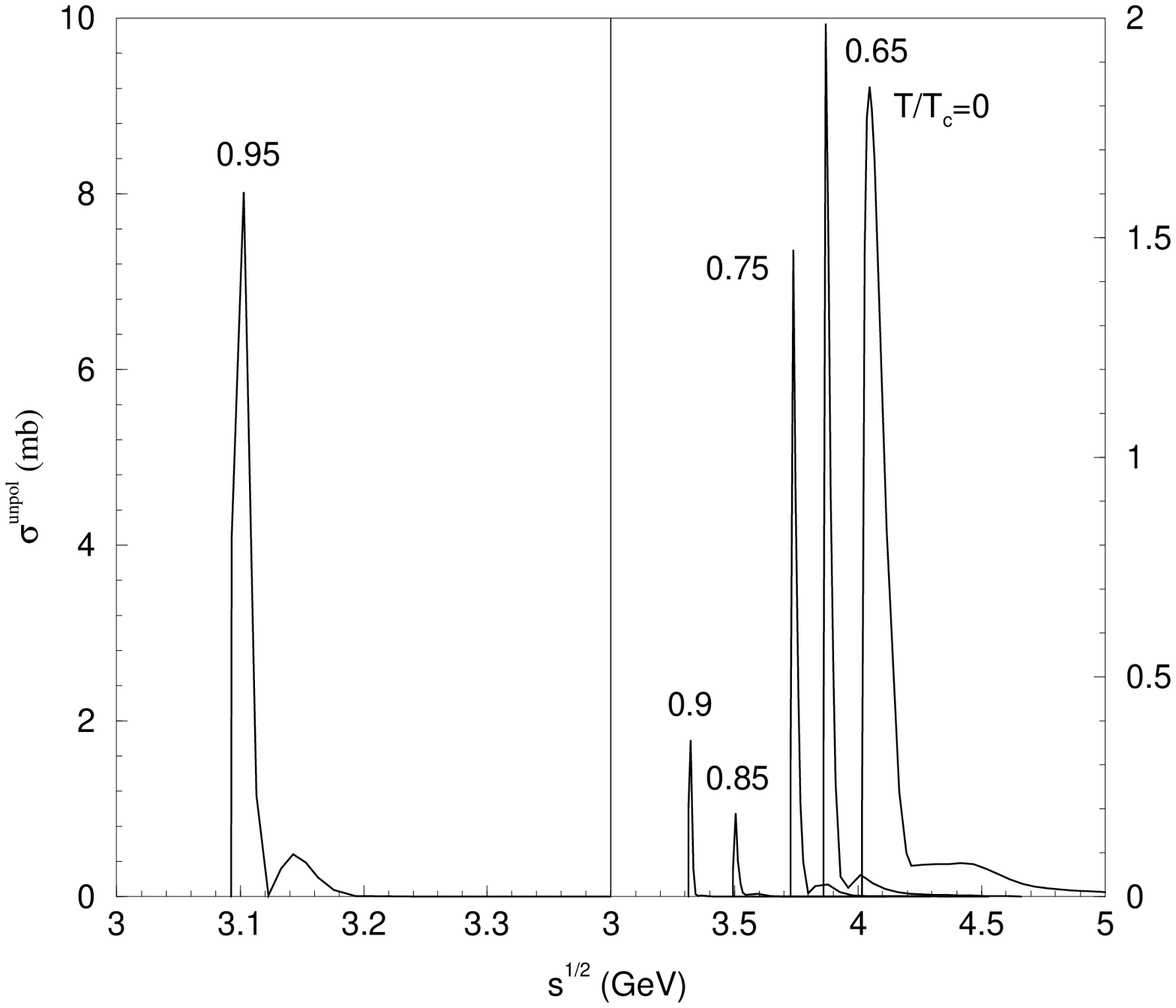}%
\caption{Cross sections for $\pi {\psi}' \to \bar{D}^* D^*$ at various 
temperatures.}
\label{fig21}
\end{figure}

\newpage

\begin{figure}[htbp]
\centering
\includegraphics[scale=0.8]{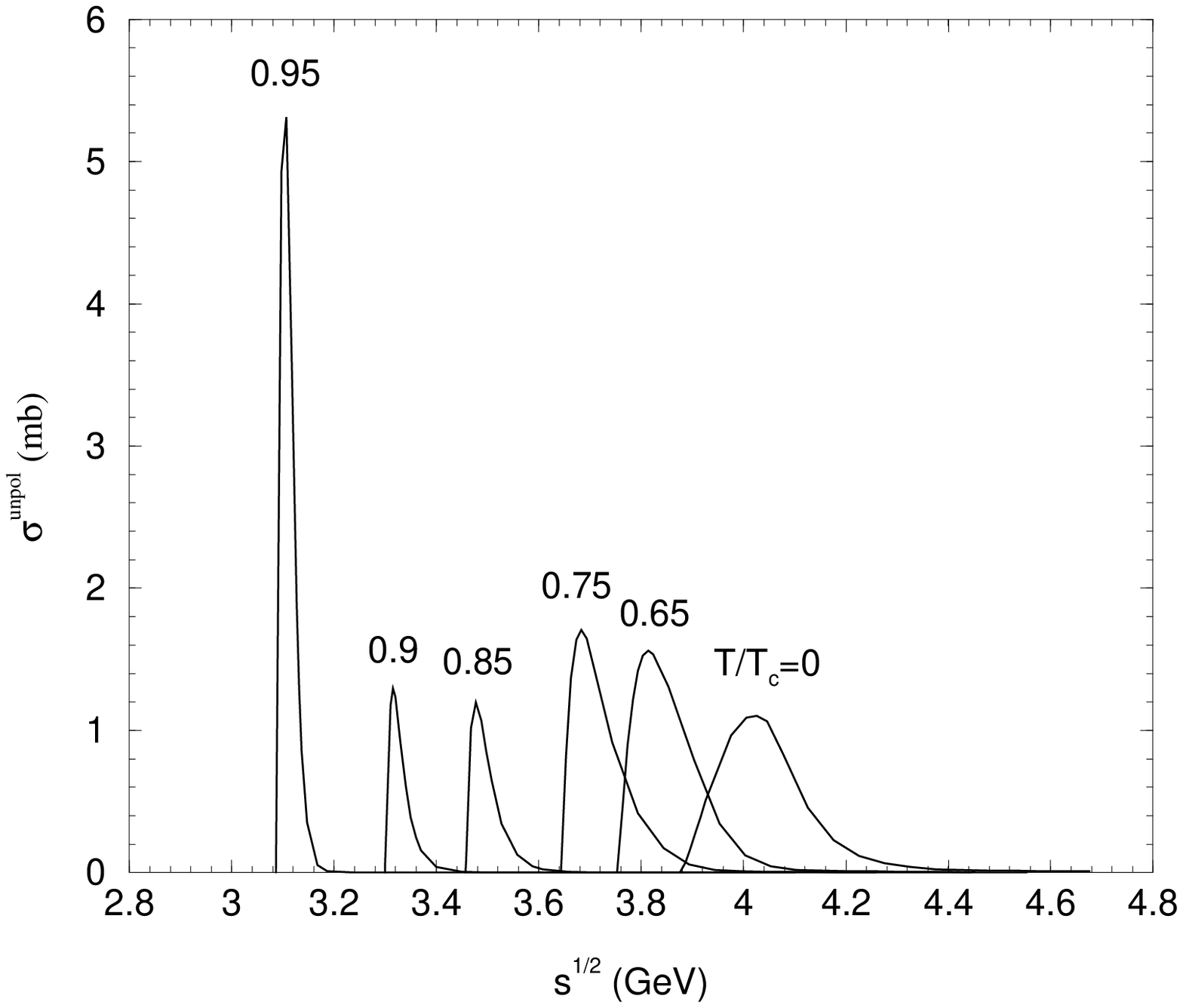}%
\caption{Cross sections for $\pi \chi_c \to \bar{D}^* D$ or $\bar D D^*$ at 
various temperatures.}
\label{fig22}
\end{figure}

\newpage

\begin{figure}[htbp]
\centering
\includegraphics[scale=0.8]{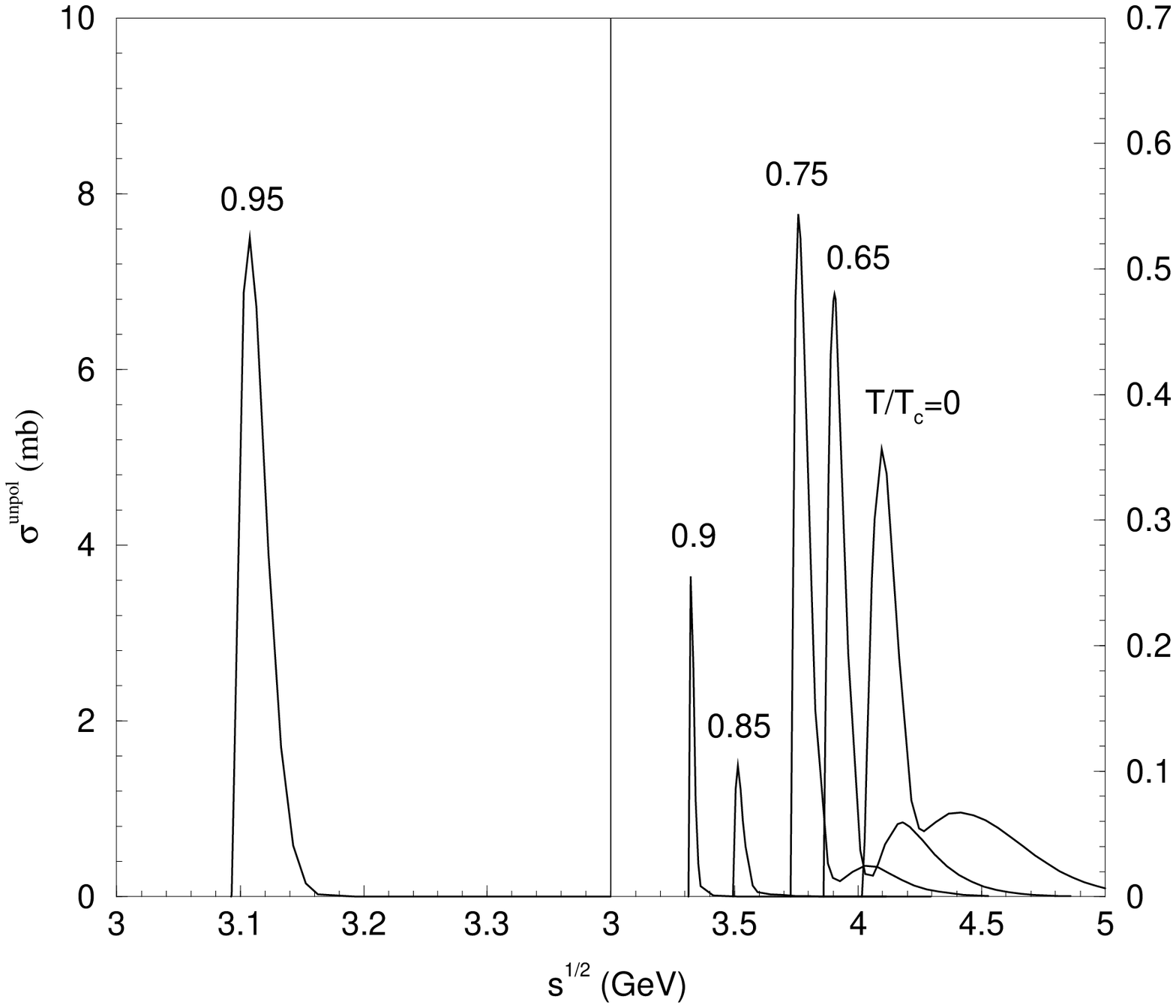}%
\caption{Cross sections for $\pi \chi_c \to \bar{D}^* D^*$ at various 
temperatures.}
\label{fig23}
\end{figure}

\newpage

\begin{table}
\centering \caption{Values of the parameters in
Eq. (30) for the $\pi J/\psi$ dissociation. $a_1$ and $a_2$ are in
units of mb; $b_1$, $b_2$, $d_0$, and $\sqrt{s_{\rm z}}$ are in units of GeV; 
$c_1$ and $c_2$ are dimensionless.}

\label{table1}
\begin{tabular*}{16cm}{@{\extracolsep{\fill}}c|c|c|c|c|c|c|c|c|c}

  \hline
  Reactions & $T/T_{\rm c}$ & $a_1$ & $b_1$ & $c_1$ & $a_2$ & $b_2$ & $c_2$ & 
$d_0$ & $\sqrt{s_{\rm z}}$ \\
  \hline
 $\pi J/\psi \to \bar{D}^* D$       & 0    & 0.16  & 0.03  & 0.53  & 0.05  
& 0.08 & 2.42 & 0.04 & 4.556 \\
                                    & 0.65 & 0.2  & 0.021  & 0.51  & 0.1  
& 0.069 & 1.8 & 0.03 & 4.279\\
 or $\bar D D^*$                    & 0.75 & 0.27  & 0.025  & 0.47  & 0.1  
& 0.044 & 0.87 & 0.03 & 4.089\\
                                    & 0.85 & 0.21  & 0.021  & 0.5  & 0.09  
& 0.073 & 1.87 & 0.03 & 3.847\\
                                    & 0.9  & 0.18  & 0.015  & 0.51  & 0.08  
& 0.069 & 1.75 & 0.02 & 3.649\\
                                    & 0.95 & 1.05  & 0.006  & 0.49  & 1  
& 0.025 & 1.27 & 0.01 & 3.291\\
  \hline
 $\pi J/\psi \to \bar{D}^* D^*$         & 0    & 0.026   & 0.26  & 4.54  
& 0.006  & 0.35 & 0.59 & 0.25 & 5.072\\
                                        & 0.65 & 0.022  & 0.2  & 2.14  
& 0.005 & 0.14 & 0.56 & 0.23 & 4.864\\
                                        & 0.75 & 0.026  & 0.2  & 3.29  
& 0.003 & 0.52 & 0.2 & 0.22 & 4.666\\
                                        & 0.85 & 0.01  & 0.016  & 0.5  
& 0.006  & 0.27 & 5.09 & 0.02 & 4.287\\
                                        & 0.9  & 0.006  & 0.039  & 1  
& 0.003  & 0.005 & 0.45 & 0.03 & 4.066\\
                                        & 0.95 & 1.61  & 0.007  & 0.54  
& 0.45 & 0.026 & 2.78 & 0.01 & 3.187\\
  \hline
\end{tabular*}
\end{table}

\newpage

\begin{table}
\centering \caption{The same as Table 1 except for $\pi \psi'$.}
\label{table2}
\begin{tabular*}{16cm}{@{\extracolsep{\fill}}c|c|c|c|c|c|c|c|c|c}

  \hline
  Reactions & $T/T_{\rm c}$ & $a_1$ & $b_1$ & $c_1$ & $a_2$ & $b_2$ & $c_2$ & 
$d_0$ & $\sqrt{s_{\rm z}}$ \\
  \hline
 $\pi{\psi}' \to \bar{D}^* D$       & 0    & 1.78  & 0.02   & 0.5  & 1.15  
& 0.06  & 1.69 & 0.035 & 4.567\\
                                    & 0.65 & 2.89  & 0.01  & 0.68   & 1.37  
& 0.01  & 0.34 & 0.01 & 4.021\\
 or $\bar D D^*$                    & 0.75 & 3.18  & 0.01   & 1.11  & 0.61  
& 0.01 & 0.01 & 0.01 & 3.871\\
                                    & 0.85 & 1.46  & 0.005  & 0.49  & 0.89   
& 0.009 & 0.55 & 0.01 & 3.606\\
                                    & 0.9  & 1.38  & 0.01  & 5.59  & 0.68  
& 0.03 & 0.01 & 0.01 & 3.431\\
                                    & 0.95 & 5.04  & 0.002  & 0.32  & 3.34  
& 0.01 & 6.61 & 0.01 & 3.215\\
  \hline
 $\pi {\psi}' \to \bar{D}^* {D}^*$      & 0    & 1.6   & 0.025   & 0.5  & 0.3
  & 0.042  & 1.66 & 0.03 & 4.78\\
                                        & 0.65 & 1.6   & 0.005  & 0.51  & 0.78
  & 0.018  & 1.64 & 0.01 & 4.103\\
                                        & 0.75 & 1.3    & 0.005  & 0.49  & 0.37
  & 0.014 & 1.42 & 0.01 & 3.916\\
                                        & 0.85 & 0.13   & 0.008  & 2.36  & 0.08
  & 0.003 & 0.12 & 0.01 & 3.64\\
                                        & 0.9  & 0.35   & 0.005  & 2.23  & 0.36
  & 0.002 & 0.28 & 0.01 & 3.349\\
                                        & 0.95 & 6.4    & 0.009  & 5.08  & 5
  & 0.001 & 0.15 & 0.01 & 3.176\\
  \hline

\end{tabular*}
\end{table}

\newpage

\begin{table}
\centering \caption{The same as Table 1 except for $\pi \chi_c$.}
\label{table3}
\begin{tabular*}{16cm}{@{\extracolsep{\fill}}c|c|c|c|c|c|c|c|c|c}

  \hline
  Reactions & $T/T_{\rm c}$ & $a_1$ & $b_1$ & $c_1$ & $a_2$ & $b_2$ & $c_2$ & 
$d_0$ & $\sqrt{s_{\rm z}}$ \\
  \hline
 $\pi\chi_c \to \bar{D}^*D$         & 0    & 0.95  & 0.13   & 3.17  & 0.18  
& 0.07  & 0.84 & 0.15 & 4.577\\
                                    & 0.65 & 1.13  & 0.07  & 1.78  & 0.5  
& 0.04  & 0.71 & 0.06 & 4.135\\
 or $\bar D D^*$                    & 0.75 & 1.39  & 0.04  & 1.32  & 0.32  
& 0.05  & 0.46 & 0.04 & 3.955\\
                                    & 0.85 & 1.04  & 0.02   & 1.04  & 0.16  
& 0.01  & 0.25 & 0.02 & 3.636\\
                                    & 0.9  & 1.14   & 0.01  & 0.83  & 0.4  
& 0.03  & 3.67 & 0.015 & 3.438\\
                                    & 0.95 & 4.51  & 0.017  & 2.03  & 1.32  
& 0.01  & 1.05 & 0.02 & 3.167\\
  \hline
 $\pi\chi_c \to \bar{D}^*{D}^*$        & 0    & 0.29   & 0.08   & 2.53  & 0.08
& 0.05  & 1.12 & 0.08 & 5.069\\
                                       & 0.65 & 0.39   & 0.04   & 1.45   & 0.1 
& 0.05  & 3.86  & 0.045 & 4.58\\
                                       & 0.75 & 0.45   & 0.03  & 1.6   & 0.11
& 0.05  & 1.17 & 0.03 & 4.293\\
                                       & 0.85 & 0.08   & 0.016  & 1.48  & 0.03 
& 0.028 & 1.65 & 0.02 & 3.724\\
                                       & 0.9  & 0.17   & 0.003  & 2.22  & 0.25
& 0.012 & 1.98 & 0.01 & 3.405\\
                                       & 0.95 & 6.36   & 0.003  & 2.43  & 7.48
& 0.015 & 2.16 & 0.015 & 3.159\\
  \hline
\end{tabular*}
\end{table}

\newpage

\begin{table}
\centering \caption{Values of the parameters in Eqs. (30) and (32) for the 
$\rho J/\psi$ dissociation. $a_1$ and $a_2$ are in
units of mb; $b_1$, $b_2$, $d_0$, and $\sqrt {s_{\rm z}}$ are in units of GeV; 
$c_1$ and $c_2$ are dimensionless.}
\label{table4}
\begin{tabular*}{16cm}{@{\extracolsep{\fill}}c|c|c|c|c|c|c|c|c|c}

  \hline
  Reactions & $T/T_{\rm c}$ & $a_1$ & $b_1$ & $c_1$ & $a_2$ & $b_2$ & $c_2$ & 
$d_0$ & $\sqrt{s_{\rm z}}$ \\
  \hline
 $\rho J/\psi \to \bar{D} D$      & 0    & 0.073   & 0.04   & 0.51  & 0.049   
& 0.45  & 5.01 & 0.05 & 4.747\\
                                  & 0.65 & 0.24   & 0.008  & 0.47  & 0.16   
& 0.03  & 0.79 & 0.01 & 4.29\\
                                  & 0.75 & 0.22   & 0.01   & 0.41  & 0.11    
& 0.01  & 0.54 & 0.007 & 3.676\\
                                  & 0.85 & 0.16   & 0.007  & 0.52  & 0.049  
& 0.02  & 0.45 & 0.007 & 3.887\\
                                  & 0.9  & 0.25   & 0.005  & 0.37  & 0.15  
& 0.01  & 1.99 & 0.01 & 3.423\\
                                  & 0.95 & 0.83    & 0.005  & 0.5   & 0.7   
& 0.02  & 1.57 & 0.01 & 3.179\\
  \hline
 $\rho J/\psi \to \bar{D}^* {D}$      & 0    & 0.71  & 0.03  & 0.39 & 0.56  
& 0.42  & 6.42 & 0.03 & 5.011\\
                                      & 0.65 & 0.28  & 0.013 & 0.5  & 0.067  
& 0.3  & 2.63 & 0.02 & 4.648\\
 or $\bar D D^*$                      & 0.75 & 0.15  & 0.012 & 0.47 & 0.044 
& 0.27  & 3.06 & 0.01 & 4.541\\
                                      & 0.85 & 0.091 & 0.009 & 0.44 & 0.045 
& 0.18  & 3.92 & 0.01 & 4.196\\
                                      & 0.9  & 0.28  & 0.005 & 0.38 & 0.15 
& 0.006 & 1.49 & 0.01 & 3.6\\
                                      & 0.95 & 2.03  & 0.007 & 0.53 & 0.68   
& 0.024 & 2.18 & 0.01 & 3.181\\
  \hline
 $\rho J/\psi \to \bar{D}^* {D}^*$     & 0    & 0.63  & 0.05  & 0.47 & 0.43  
& 0.36  & 5.79 & 0.04 & 5.043\\
                                       & 0.65 & 0.086 & 0.06  & 0.46 & 0.043  
& 0.28  & 4.05 & 0.04 & 4.901\\
                                       & 0.75 & 0.043 & 0.03  & 0.46 & 0.033 
& 0.24  & 2.39 & 0.03 & 4.765\\
                                       & 0.85 & 0.0061 & 0.012 & 0.46 & 0.0031 
& 0.14  & 1.6 & 0.01 & 4.291\\
                                       & 0.9  & 0.27  & 0.007 & 0.84 & 0.091 
& 0.002 & 0.18 & 0.01 & 3.607\\
                                       & 0.95 & 5.51  & 0.005 & 0.51 & 4.47   
& 0.02  & 1.63 & 0.01 & 3.189\\
  \hline
 \end{tabular*}
\end{table}

\newpage

\begin{table}
\centering \caption{The same as Table 4 except for $\rho {\psi}'$.}
\label{table5}
\begin{tabular*}{16cm}{@{\extracolsep{\fill}}c|c|c|c|c|c|c|c|c|c}

  \hline
  Reactions & $T/T_{\rm c}$ & $a_1$ & $b_1$ & $c_1$ & $a_2$ & $b_2$ & $c_2$ & 
$d_0$ & $\sqrt{s_{\rm z}}$\\
  \hline
 $\rho {\psi}' \to \bar{D} D$      & 0    & 0.02   & 0.13  & 0.72  & 0.02   
& 0.05  & 0.45 & 0.1 & 5.355\\
                                   & 0.65 & 0.0019 & 0.04  & 0.53  & 0.0012  
& 0.25  & 2.74 & 0.03 & 4.848\\
                                   & 0.75 & 0.0022 & 0.022 & 0.54  & 0.00079 
& 0.27  & 4.83 & 0.03 & 4.37\\
                                   & 0.85 & 0.049  & 0.01  & 0.52  & 0.047  
& 0.021 & 2.76 & 0.01 & 3.575\\
                                   & 0.9  & 3.1    & 0.006 & 0.9   & 5.57    
& 0.002 & 0.38 & 0.01 & 3.373\\
                                   & 0.95 & 1.07   & 0.005 & 0.62  & 0.26   
& 0.01  & 0.1 & 0.01 & 3.171\\
  \hline
 $\rho {\psi}' \to \bar{D}^* {D}$         & 0    & 0.083 & 0.05   & 0.5  
& 0.045   & 0.23   & 1.8 & 0.1 & 5.354\\
                                          & 0.65 & 0.017 & 0.03   & 0.53 
& 0.0043 & 0.3    & 8.68 & 0.03 & 4.818\\
 or $\bar D D^*$                          & 0.75 & 0.027  & 0.008  & 0.46 
& 0.021  & 0.04   & 0.74 & 0.01 & 4.44\\
                                          & 0.85 & 3.6   & 0.0008 & 0.49 
& 4.54   & 0.0046 & 1.27 & 0.005 & 3.568\\
                                          & 0.9  & 2.09  & 0.006  & 1.8  
& 2.09   & 0.003  & 0.17 & 0.01 & 3.385\\
                                          & 0.95 & 0.99  & 0.006  & 0.77 
& 0.62   & 0.007  & 0.23 & 0.01 & 3.173\\
  \hline
 $\rho {\psi}'\to \bar{D}^* {D}^*$        & 0    & 0.2  & 0.03   & 0.51  
& 0.14 & 0.19   & 1.37 & 0.05 & 5.337\\
                                          & 0.65 & 0.78 & 0.05   & 3.08  
& 0.04 & 0.0013 & 0.68 & 0.05 & 4.535\\
                                          & 0.75 & 10.86  & 0.0009 & 0.44  
& 13.39  & 0.005  & 1.05 & 0.005 & 3.832\\
                                          & 0.85 & 4.79   & 0.005  & 0.82  
& 6.23 & 0.002  & 0.48 & 0.01 & 3.597\\
                                          & 0.9  & 2.07 & 0.003  & 0.49  
& 0.72 & 0.012  & 0.78 & 0.004 & 3.394\\
                                          & 0.95 & 2.76 & 0.005  & 0.47  
& 1.08 & 0.015  & 0.8 & 0.006 & 3.185\\
  \hline
\end{tabular*}
\end{table}

\newpage

\begin{table}
\centering \caption{The same as Table 4 except for $\rho \chi_c$.}
\label{table6}
\begin{tabular*}{16cm}{@{\extracolsep{\fill}}c|c|c|c|c|c|c|c|c|c}

  \hline
  Reactions & $T/T_{\rm c}$ & $a_1$ & $b_1$ & $c_1$ & $a_2$ & $b_2$ & $c_2$ & 
$d_0$ & $\sqrt{s_{\rm z}}$ \\
  \hline
 $\rho \chi_c \to \bar{D} D$      & 0    & 0.029  & 0.07  & 0.5  & 0.0073   
& 0.21  & 1.24 & 0.1 & 5.173\\
                                  & 0.65 & 0.007  & 0.21  & 3.1  & 0.0003 
& 0.005 & 0.51 & 0.2 & 4.815\\
                                  & 0.75 & 0.001  & 0.007 & 0.51 & 0.0008 
& 0.24  & 3.59 & 0.01 & 4.306\\
                                  & 0.85 & 0.03   & 0.003 & 0.63 & 0.0062  
& 0.01  & 0.2 & 0.005 & 3.666\\
                                  & 0.9  & 2.98   & 0.007 & 1.42 & 1.94   
& 0.017 & 2.98 & 0.01 & 3.397\\
                                  & 0.95 & 0.52   & 0.009 & 1.54 & 0.32   
& 0.019 & 3.15 & 0.01 & 3.203\\
  \hline
 $\rho \chi_c \to \bar{D}^* {D}$      & 0    & 0.1    & 0.16  & 1.16 & 0.045   
& 0.02  & 0.49 & 0.15 & 5.172\\
                                      & 0.65 & 0.0029 & 0.008 & 0.44 & 0.018  
& 0.23  & 2.7 & 0.25 & 4.806\\
 or $\bar D D^*$                      & 0.75 & 0.0034 & 0.007 & 0.5  & 0.0044 
& 0.2   & 3.09 & 0.15 & 4.302\\
                                      & 0.85 & 1.75   & 0.003 & 0.5  & 1.41   
& 0.011 & 1.63 & 0.005 & 3.606\\
                                      & 0.9  & 1.58   & 0.01  & 1.4  & 0.81   
& 0.015 & 2.03 & 0.01 & 3.433\\
                                      & 0.95 & 0.34   & 0.014 & 1.33 & 0.11   
& 0.013 & 4.84 & 0.01 & 3.222\\
  \hline
 $\rho \chi_c \to \bar{D}^* {D}^*$       & 0    & 0.31 & 0.27  & 3.43 & 0.29 
& 0.04  & 0.54 & 0.25 & 5.172\\
                                         & 0.65 & 1.91 & 0.005 & 0.52 & 1.87  
& 0.024 & 1.55 & 0.01 & 4.44\\
                                         & 0.75 & 8.57 & 0.006 & 0.58 & 8.57   
& 0.018 & 1.82 & 0.01 & 3.879\\
                                         & 0.85 & 2.21 & 0.007 & 1.58 & 1.96  
& 0.018 & 3.48 & 0.01 & 3.632\\
                                         & 0.9  & 0.29 & 0.012 & 1.66 & 0.13 
& 0.013 & 1.02 & 0.01 & 3.459\\
                                         & 0.95 & 0.11 & 0.013 & 1.99 & 0.041 
& 0.025 & 0.76 & 0.01 & 3.231\\
  \hline
\end{tabular*}
\end{table}


\newpage

\begin{thebibliography}{99}
\addtolength{\itemsep}{-0.6 em}
\bibitem{MS}T. Matsui and H. Satz, Phys. Lett. B 178, 416 (1986).
\bibitem{AT}J. Schukraft, Talk presented at the 22nd International Conference 
on Ultra-Relativistic Nucleus-Nucleus Collisions, Annecy, France, May 23-28, 
2011.
\bibitem{BW}B. Wyslouch, Talk presented at the 22nd International Conference on
Ultra-Relativistic Nucleus-Nucleus Collisions, Annecy, France, May 23-28, 2011.
\bibitem{FLP}J. Ft$\acute{\rm a}\check{\rm c}$nik, P. Lichard, and J. 
Pi$\check{\rm s}\acute{\rm u}$t, Phys. Lett. B 207, 194 (1988); S. Gavin, M. 
Gyulassy, and A. Jackson, Phys. Lett. B 207, 257 (1988); R. Vogt, M. Prakash, 
P. Koch, and T. H. Hansson, Phys. Lett. B 207, 263 (1988); C. Gerschel and J. 
H\"ufner, Phys. Lett. B 207, 253 (1988).
\bibitem{MEP}M. E. Peskin, Nucl. Phys. B 156, 365 (1979); G. Bhanot and M. E. 
Peskin, Nucl. Phys. B 156, 391 (1979).
\bibitem{KS}D. Kharzeev and H. Satz, Phys. Lett. B 334, 155 (1994).
\bibitem{AGGA}F. Arleo, P. B. Gossiaux, T. Gousset, and J. Aichelin, Phys. Rev.
D 65, 014005 (2001).
\bibitem{MM}S. G. Matinyan and B. M\"uller, Phys. Rev. C 58, 2994 (1998).
\bibitem{LK}Z. Lin and C. M. Ko, Phys. Rev. C 62, 034903 (2000); J. Phys. G 27,
617 (2001).
\bibitem{KLH}K. L. Haglin, Phys. Rev. C 61, 031902 (2000); K. L. Haglin and C. 
Gale, Phys. Rev. C 63, 065201 (2001).
\bibitem{OSL}Y. Oh, T. Song, and S. H. Lee, Phys. Rev. C 63, 034901 (2001).
\bibitem{NNR}F. S. Navarra, M. Nielsen, and M. R. Robilotta, Phys. Rev. C 64, 
021901(R) (2001).
\bibitem{MPPR}L. Maiani, F. Piccinini, A. D. Polosa, and V. Riquer, Nucl. Phys.
A 741, 273 (2004).
\bibitem{BG}A. Bourque and C. Gale. Phys. Rev. C 78, 035206 (2008); Phys. Rev. 
C 80, 015204 (2009).
\bibitem{MBQ}K. Martins, D. Blaschke, and E. Quack, Phys. Rev. C 51, 2723 
(1995).
\bibitem{WSB}C.-Y. Wong, E. S. Swanson, and T. Barnes, Phys. Rev. C 62, 045201 
(2000).
\bibitem{WSB01}C.-Y. Wong, E. S. Swanson, and T. Barnes, Phys. Rev. C 65, 
014903 (2001).
\bibitem{BSWX}T. Barnes, E. S. Swanson, C.-Y. Wong, and X.-M. Xu, Phys. Rev. C 
68, 014903 (2003).
\bibitem{BS92}T. Barnes and E. S. Swanson, Phys. Rev. D 46, 131 (1992); E. S. 
Swanson, Ann. Phys. (N.Y.) 220, 73 (1992).
\bibitem{ZXG}Y.-P. Zhang, X.-M. Xu, and H.-J. Ge, Nucl. Phys. A 832, 112 
(2010).
\bibitem{DGG}A. De R$\acute{\rm u}$jula, H. Georgi, and S. L. Glashow, Phys. 
Rev. D 12, 147 (1975).
\bibitem{IS1}N. Isgur and G. Karl, Phys. Rev. D 18, 4187 (1978); Phys. Rev. D 
19, 2653 (1979); Phys. Rev. D 20, 1191 (1979).
\bibitem{GI}S. Godfrey and N. Isgur, Phys. Rev. D 32, 189 (1985).
\bibitem{CI}S. Capstick and N. Isgur, Phys. Rev. D 34, 2809 (1986).
\bibitem{KLP01}F. Karsch, E. Laermann, and A. Peikert, Nucl. Phys. B 605, 579 
(2001).
\bibitem{BT}W. Buchm\"{u}ller and S.-H. H. Tye, Phys. Rev. D 24, 132 (1981).
\bibitem{EPRD}E. Eichten $\it {et}$ $\it {al.}$, Phys. Rev. D 17, 3090 (1978);
 Phys. Rev. D 21, 203 (1980).
\bibitem{KN2010}K. Nakamura $\it {et}$ $\it {al.}$ (Particle Data Group), J. 
Phys. G 37, 075021 (2010).
\bibitem{XMX}X.-M. Xu, Nucl. Phys. A 697, 825 (2002).
\bibitem{LX}Y.-Q. Li and X.-M. Xu, Nucl. Phys. A 794, 210 (2007).
\bibitem{MM65}N. F. Mott and H. S. W. Massey, The Theory of Atomic Collisions, 
Clarendon Press, Oxford, 1965.
\bibitem{BBS01}T. Barnes, N. Black, and E. S. Swanson, Phys. Rev. C 63, 025204 
(2001).
\bibitem{WC}C.-Y. Wong and H. W. Crater, Phys. Rev. C 63, 044907 (2001).
\bibitem{EC}E. Colton $\it {et}$ $\it {al.}$, Phys. Rev. D 3, 2028 (1971).
\bibitem{NBD}N. B. Durusoy $\it {et}$ $\it {al.}$, Phys. Lett. B 45, 517 
(1973).
\bibitem{WH}W. Hoogland $\it {et}$ $\it {al.}$, Nucl. Phys. B 126, 109 (1977).
\bibitem{MJL}M. J. Losty $\it {et}$ $\it {al.}$, Nucl. Phys. B 69, 185 (1974).
\bibitem{Wong}C.-Y. Wong, Phys. Rev. C 65, 034902 (2002).
\end{thebibliography}
\end{document}